\documentclass[letterpaper,titlepage,11pt]{article}

\pdfoutput=1

\usepackage{amssymb,amsmath,amsfonts}
\usepackage{epsfig}
\usepackage{ccaption}
\usepackage{graphicx}
\usepackage{subfig}

\usepackage[
      colorlinks=true,
      linkcolor=blue,
      urlcolor=blue,
      filecolor=black,
      citecolor=red,
      pdfstartview=FitV,
      pdftitle={Kerr Merger},
        pdfauthor={Roberto Emparan, Marina Martinez, Miguel Zilhão},
        pdfsubject={},
        pdfkeywords={},
        pdfpagemode=None,
        bookmarksopen=true
      ]{hyperref}

\def\clock{{\count0=\time
           \divide\count0 60
           \ifnum\count0<10 0\fi\the\count0
           \multiply\count0 -60 \advance\count0 \time
           :\ifnum\count0<10 0\fi \the\count0
         }}
\newcommand{\timestamp}{{\small\vbox{\hbox{\tt\jobname.tex}
\hbox{\the\day/\the\month/\the\year, \clock}}}}

\newcommand{\nn}{\nonumber}
\newcommand{\ie}{{\it i.e.,\,}}
\newcommand{\eg}{{\it e.g.,\,}}

\newcommand{\lp}{\left(}
\newcommand{\rp}{\right)}

\newcommand{\beq}{\begin{equation}}
\newcommand{\eeq}{\end{equation}}
\newcommand{\bea}{\begin{eqnarray}}
\newcommand{\eea}{\end{eqnarray}}
\newcommand{\beqa}{\begin{eqnarray}}
\newcommand{\eeqa}{\end{eqnarray}}

\numberwithin{equation}{section}

\begin{document}

\begin{titlepage}
\leftline{}
\vskip 2cm
\centerline{\LARGE \bf Black hole fusion} 
\bigskip
\centerline{\LARGE \bf in the extreme mass ratio limit}
\vskip 1.2cm
\centerline{\bf Roberto Emparan$^{a,b}$, Marina Mart{\'\i}nez$^{b}$, Miguel Zilh\~ao$^{b,c}$}
\vskip 0.5cm
\centerline{\sl $^{a}$Instituci\'o Catalana de Recerca i Estudis
Avan\c cats (ICREA)}
\centerline{\sl Passeig Llu\'{\i}s Companys 23, E-08010 Barcelona, Spain}
\smallskip
\centerline{\sl $^{b}$Departament de F{\'\i}sica Qu\`antica i Astrof\'{\i}sica, Institut de
Ci\`encies del Cosmos,}
\centerline{\sl  Universitat de
Barcelona, Mart\'{\i} i Franqu\`es 1, E-08028 Barcelona, Spain}
\smallskip
\centerline{\sl $^{c}$CENTRA, Departamento de F\'\i sica, Instituto Superior T\'ecnico, }
\centerline{\sl  Universidade de Lisboa, Avenida Rovisco Pais 1, 1049 Lisboa, Portugal}
\smallskip

\vskip 1.2cm
\centerline{\bf Abstract} \vskip 0.2cm 
\noindent 
We present a simple, general, and accurate construction of the event horizons for the fusion of two neutral, rotating black holes with arbitrary orientation and values of their spins, in the extreme mass ratio limit where one black hole is much larger than the other. We compute several parameters that characterize the fusion and investigate their dependence on the black hole spin and orientation axis. We also exhibit and study the appearance of transient toroidal topology of the horizon. 
An earlier conjecture about universal critical exponents before and after an axisymmetric pinch is proven. 

\end{titlepage}
\pagestyle{empty}
\small
\tableofcontents
\normalsize
\newpage
\pagestyle{plain}
\setcounter{page}{1}

\section{Introduction}

The event horizon has a central status in the theory of black holes. Several kinds of horizon have been introduced to characterize a black hole, and all of them suffer from one oddity or another---for instance, they cannot avoid some form of non-locality (which is then a feature, not a bug). However, the event horizon stands out among all of them as the one that neatly captures the core intuition that nothing, not even light, can ever escape from the black hole. Moreover, it is a very useful concept in General Relativity, since one can prove with relative ease a number of far-reaching theorems that the event horizon must satisfy, most remarkably the area law \cite{Hawking:1971tu}.

Given its importance, one is naturally interested in the shape and features of the event horizon in phenomena where black hole dynamics is in full swing. None is more compelling nor challenging than the fusion of two black holes.

Progress in identifying important properties of the event horizon of a binary black hole collision was made, following early numerical analysis in \cite{Hughes:1994ea}, using a clever, simple exact construction of a null hypersurface that models the merger horizon \cite{Shapiro:1995rr,Lehner:1998gu,Husa:1999nm}. Still, it is commonly assumed that the identification of such time-evolving horizons in an actual collision requires heavy computational resources, if only because solving the equations to obtain the complete spacetime geometry of a black hole merger is a very hard problem. However, we have recently shown that when the ratio of the two black hole masses, $m$ and $M$, is very small, the fusion event horizon can be obtained very easily, even in closed exact form when $m/M\to 0$ \cite{Emparan:2016ylg}. The reason is that in this limit, with $m$ kept finite while $M\to\infty$, the spacetime geometry around the fusion region is known, and it is simple: it is the Schwarzschild, or more generally the Kerr geometry.\footnote{Note that, even if this is indeed the extreme mass ratio limit, it is not the point particle limit for the small black hole (see \cite{Hamerly:2010cr,Hussain:2017ihw} for the event horizon in the latter case). The reason is that we intend to resolve the distance scales $\sim m$ that are involved in the fusion, instead of the lengths $\sim M$ associated to the inspiral gravitational radiation.} 

The argument is elementary. When the radius of the large black hole is sent to infinity, the curvature that it creates in any finite patch of spacetime vanishes. Then the curvature in the region where the fusion happens can only be due to the gravity of the small black hole. The spacetime here must be very well approximated (exactly when $m/M\to 0$) by the geometry of a black hole with mass $m$. Observe that this is essentially a consequence of the equivalence principle: at scales much shorter than $M$, and in its own rest frame, the small black hole is freely falling in flat space, and the large black hole horizon is just an acceleration horizon.

Since, by this reasoning, we know the spacetime geometry of the merger (within distances $\ll M$), the determination of the event horizon reduces, as always, to finding an appropriate congruence of null geodesics in it. In the case at hand, we must select a congruence that approaches a null plane at late times: think of a Rindler horizon, \ie\ the limiting shape of the infinitely large black hole after it has relaxed to equilibrium. This limit provides the final condition on the null geodesic generators of the event horizon. Then, starting from these final points we can integrate the null geodesic equations back in time in order to obtain the complete null hypersurface that they rule. Up to the identification of the \textit{crease set} of the horizon---the set of points where null rays focus (caustic points) or cross each other (crossover points) as they enter the future horizon \cite{Shapiro:1995rr,Lehner:1998gu,Husa:1999nm,Siino:2004xe}---this hypersurface is the event horizon of the merger.

The integration of geodesics can be readily performed in closed analytic form when the small black hole is not rotating, \ie\ it is a Schwarzschild black hole. The resulting event horizon was thoroughly studied in \cite{Emparan:2016ylg}. In this paper we construct it for the case in which the small black hole is rotating, and thus given by the Kerr solution. The geodesic equations in Kerr are integrable, \ie\ can be reduced to quadratures \cite{Carter:1968rr}, but even if an analytical solution for the null geodesics is still possible (see \eg\ \cite{cadez:1998}), it is too complicated to be of practical use for us here. Therefore in this paper we will integrate the geodesic equations numerically, since they are simple enough and easily reveal the main features of the event horizon.

It turns out that this construction is the ultimate calculation of the event horizon for any extreme-mass-ratio fusion of two black holes (at least astrophysically relevant ones) in the limit $m/M\to 0$. That is, our analysis covers the most general instance of these mergers: it includes any rotation of either of the two black holes, any relative velocity between them, and arbitrary alignments between the spins and the collision trajectory. 

As we will argue below, at leading order in $m/M\to 0$ there are only two non-trivial parameters that specify the event horizon of any extreme-mass-ratio fusion of neutral black holes. They are the spin parameter $a$ of the small black hole (measured in units of the mass $m$), and the angle $\alpha$ between this spin axis and the collision axis. By varying them in the ranges $0\leq a/m\leq 1$ and $0\leq \alpha\leq \pi/2$, one covers the entire spectrum of extreme-mass-ratio black hole mergers in the Universe.

\paragraph{Plan.} According to the previous discussion, in order to find the event horizon for a binary merger when $m$ is finite and $M\to\infty$ we must solve the following problem in General Relativity: 
\begin{enumerate}
\item[a.] Obtain the null geodesic equations in the Kerr spacetime. 
\item[b.] Find appropriate final conditions for them. 
\item[c.] Integrate the equations with these final conditions, and cut out the resulting surface at the crease set, where the generators enter the event horizon.
\end{enumerate}
The main physical input enters at step b, which determines the null hypersurface we seek, and at step c in the identification of the caustic and crossover points, which restrict the hypersurface so it becomes a future event horizon.

We present the geodesic equations in the next section.
Step b is explained in sec.~\ref{sec:finalconds}. The numerical integration in step c is straightforward and discussed in sec.~\ref{sec:integresults}. The properties of the event horizon and the identification of the crease set will be illustrated in specific examples in secs.~\ref{sec:alpha0}, \ref{sec:alphapi2} and \ref{sec:alphapi4}. Section~\ref{sec:locpinch} discusses the local structure of an axisymmetric event horizon around the moment when the two black hole horizons first touch, and sec.~\ref{sec:complete} presents the argument for the completeness of this construction. Finally we make some concluding remarks in sec.~\ref{sec:concl}.

\paragraph{Note:} Unless explicitly specified, all the plots presented in the paper are in units of $m=1/2$.

\section{Null geodesic equations}

We begin with the Kerr geometry in Boyer-Lindquist coordinates,
\begin{eqnarray}
ds^2&=& -\left(1-\frac{2\,m\, r}{\Sigma}\right) dt^2-\frac{4\, m\, a\, r \sin^2\theta}{\Sigma}\, dt\, d\varphi
+ \frac{\Sigma}{\Delta}\, dr^2 +\Sigma\, d\theta^2\nn
\\&&+ \left(r^2+a^2+\frac{2\, m\, a^2 r \sin^2\theta}{\Sigma}\right)\sin^2\theta\, d\varphi^2,
\end{eqnarray}
where
\begin{equation}
\Delta\equiv r^2-2\,m\,r+a^2,\qquad \Sigma\equiv r^2+a^2\cos^2\theta.
\end{equation}

There are several possible representations for the null geodesic equations, with different variables. In the present instance, what is the most convenient choice depends on how one is going to perform their integration. 

When solving this problem for the Schwarzschild spacetime in \cite{Emparan:2016ylg}, the integrations were done in closed analytic form. For this purpose the usual Lagrangian formulation, which gives equations for $x^\mu(\lambda)$, was good enough. However, for reasons that we will discuss presently, this is not so convenient for the numerical integrations that we intend to perform here. Instead, we will work in Hamiltonian formalism, with canonical momenta $p_\mu(\lambda)$ in addition to the positions $x^\mu(\lambda)$. 

The equations were already derived in this manner in \cite{Carter:1968rr}. One naturally uses the existing constants of motion along the geodesics, namely the energy and angular momentum
\beq\label{eq:EL}
p_t=E\,, \qquad p_\varphi=L\,,
\eeq
and Carter's constant $Q$. Then, the equations for the null geodesics can be written as 
\begingroup
\allowdisplaybreaks
\begin{subequations}
\label{eq:geo}
\begin{align}
  \dot{t} &= \frac{1}{2\, \Delta\, \Sigma} \frac{\partial}{\partial E} \left( R + \Delta \Theta \right) \label{eq:tdot} \\
  \dot{r} &= \frac{\Delta}{\Sigma}\,p_{r}  \label{eq:rdot} \\
  \dot{\theta} &= \frac{1}{\Sigma}\,p_{\theta}  \label{eq:thetadot} \\
  \dot{\varphi} &= -\frac{1}{2\, \Delta\, \Sigma} \frac{\partial}{\partial L} \left( R + \Delta \Theta \right)  \label{eq:phidot} \\
  \dot{p_{t}} &= 0 \\
  \dot{p_{r}} &= - \frac{\partial}{\partial r} \left( \frac{\Delta}{2\, \Sigma}\right) p_{r}^{2}
                - \frac{\partial}{\partial r} \left( \frac{1}{2\, \Sigma}\right) p_{\theta}^{2}
                + \frac{\partial}{\partial r}
                \left( \frac{R+\Delta \Theta}{2 \, \Delta\, \Sigma} \right) \\
  \dot{p_{\theta}} &= -\frac{\partial}{\partial \theta}
                     \left( \frac{\Delta}{2\, \Sigma}\right) p_{r}^{2}
                     - \frac{\partial}{\partial \theta}
                     \left( \frac{1}{2\, \Sigma}\right) p_{\theta}^{2}
                     + \frac{\partial}{\partial \theta} \left( \frac{R+\Delta \Theta}
                     {2 \, \Delta\, \Sigma} \right) \\
\dot{p_{\varphi}} &= 0, 
\end{align}
\end{subequations}
\endgroup
(see \eg\ \cite{Fuerst:2004ii,Pu:2016eml}) with
\begin{align}
\Theta &\equiv Q+ \cos^2\theta \left(a^2 E^2 - \frac{L^2}{\sin^2\theta} \right) \\
R&\equiv \left(E(r^2+a^2)-a\, L\right)^2 - \Delta\left(Q+(L-a\,E)^2\right).
\end{align}
Dots indicate derivatives with respect to the affine parameter $\lambda$ along the geodesics. 

Carter's constant is directly related to the possibility of separating the variables $r$ and $\theta$,\footnote{The separation constant is $Q+(L-aE)^2$.} which allows to solve for the non-conserved canonical momenta as
\beq
p_r=\frac{\sqrt{R}}{\Delta}\,,\qquad p_\theta=\sigma\sqrt{\Theta}\, \quad (\sigma=\pm 1)\,.
\label{eq:prptheta}
\eeq
We may now plug these into \eqref{eq:rdot} and \eqref{eq:thetadot} to obtain
\beq\label{eq:dotrdottheta}
\dot{r}=\frac{\sqrt{R}}{\Sigma}\,,\qquad \dot{\theta}=\sigma\frac{\sqrt{\Theta}}{\Sigma}\,.
\eeq
These two equations form a closed system that (when final conditions are specified) can be solved to yield $r(\lambda)$ and $\theta(\lambda)$. Having these, the solutions for $t(\lambda)$ and $\varphi(\lambda)$ are found by quadratures. However, the exact analytic solution to equations \eqref{eq:dotrdottheta} is too unwieldy for our purposes. On the other hand, their numerical integration suffers from the problem that the surds in \eqref{eq:dotrdottheta} change sign along the geodesics, \eg\ at minima of the radial coordinate. Tracking these minima numerically is cumbersome.

Our strategy is to use \eqref{eq:dotrdottheta} to obtain a perturbative series solution in the late-time asymptotic limit $\lambda\to\infty$ which incorporates the appropriate final conditions. We will then continue numerically the integration of the geodesics back in time, but now solving the surd-free equations \eqref{eq:geo}. Let us mention that the latter were used in \cite{Fuerst:2004ii,Pu:2016eml} for this same reason.

\section{The final cut: late-time asymptotics of the event horizon}\label{sec:finalconds}

The geodesic equations \eqref{eq:geo} are first order equations that need to be supplemented with integration constants: final conditions that specify a three-dimensional null congruence that asymptotes to a given null plane. The latter corresponds to the large black hole horizon after it has relaxed back to equilibrium. 
Once these constants are appropriately fixed, the equations can be integrated to obtain the null generators. The null hypersurface that leaps out of this calculation is the event horizon, up to the excision at caustics and crossovers.

\subsection{Fixing the integration constants}\label{subsec:fixconst}

The momenta $p_t$ and $p_\varphi$ have already been integrated in \eqref{eq:EL}. For null trajectories, after the derivative $\partial/\partial E$ in \eqref{eq:tdot} is taken, we can always normalize the affine parameter $\lambda$ so as to set 
\beq
E=1\,.
\eeq

Let us now expand the equations \eqref{eq:tdot}, \eqref{eq:phidot}  and \eqref{eq:dotrdottheta}, for $\dot{t}$, $\dot{r}$, $\dot{\theta}$ and $\dot{\varphi}$ at large distance $r\gg m, a$ and large values of the affine parameter $\lambda$, and integrate them to find
\begingroup
\begin{subequations}
\label{eq:consts}
\begin{align}
t(\lambda)&= \lambda + t_\infty+2m \log\lambda +\mathcal{O}(1/\lambda)\,,\label{eq:tinf}\\
r(\lambda)&=\lambda +r_\infty+\mathcal{O}(1/\lambda)\,,\\
\theta(\lambda)&=\theta_\infty+\mathcal{O}(1/\lambda)\,,\\
\varphi(\lambda)&=\varphi_\infty+\mathcal{O}(1/\lambda)\,.
\end{align}
\end{subequations}
\endgroup
We can immediately fix at will three of the integration constants as
\beq
t_\infty=0\,,\qquad \varphi_\infty=0\,,\qquad r_\infty=0\,.
\eeq
For the first two, the isometries of the geometry guarantee that we can shift the retarded time $t-r$ of the asymptotic null plane and its orientation along the $\varphi$-circles without any loss of generality. The value of $r_\infty$ is arbitrarily chosen using the shift reparametrization symmetry of $\lambda$.

The integration constant $\theta_\infty$ corresponds to the angle between the rotation axis of the Kerr black hole and the normal to the null plane at infinity. This is a physically meaningful parameter of the merger. We will denote it as
\beq
\theta_\infty=\alpha\,.
\eeq

The asymptotic solution for $p_r$ and $p_{\theta}$ is now completely determined from \eqref{eq:prptheta} as
\beqa
p_r&=&1+\mathcal{O}(1/\lambda)\,,\qquad p_{\theta}= P_\alpha +\mathcal{O}(1/\lambda)\,,
\eeqa
where for convenience we have defined
\beq\label{Palpha}
P_\alpha=\sigma\sqrt{Q+a^2\cos^2\alpha-L^2\cot^2\alpha}\,.
\eeq
This is a quantity that is obviously conserved for geodesics in any configuration with fixed $\alpha$ and $a$. One advantage of using it instead of $Q$ is that it incorporates information about the angle $\alpha$, which $Q$ does not. Moreover, the sign of $P_\alpha$ now absorbs $\sigma$, the relative sign between $\dot r$ and $\dot{\theta}$ at large radii. This sign tells us whether a geodesic that reaches $\theta\to\alpha$ at infinity arrived there by passing above ($\sigma=+1$) or below ($\sigma=-1$) the small black hole.

\medskip

At this stage all the integration constants in the problem have been fixed. We conclude that the only physical parameters of the configuration are the angle $\alpha$ and the dimensionless rotation parameter $a/m$.

One can now proceed to integrate the equations perturbatively to higher orders in $1/\lambda$. Although this perturbative solution cannot resolve the merger region where $r\sim m$, it is useful for providing accurate initial (\ie final) conditions to perform numerical integrations that begin at a large but necessarily finite value of $\lambda$. The results of this expansion are given in appendix \ref{app:asympsol}.

\subsection{$L$, $Q$, $P_\alpha$ on the asymptotic null plane}

\begin{figure}[t]
	\centering
	\includegraphics[width=0.6\textwidth]{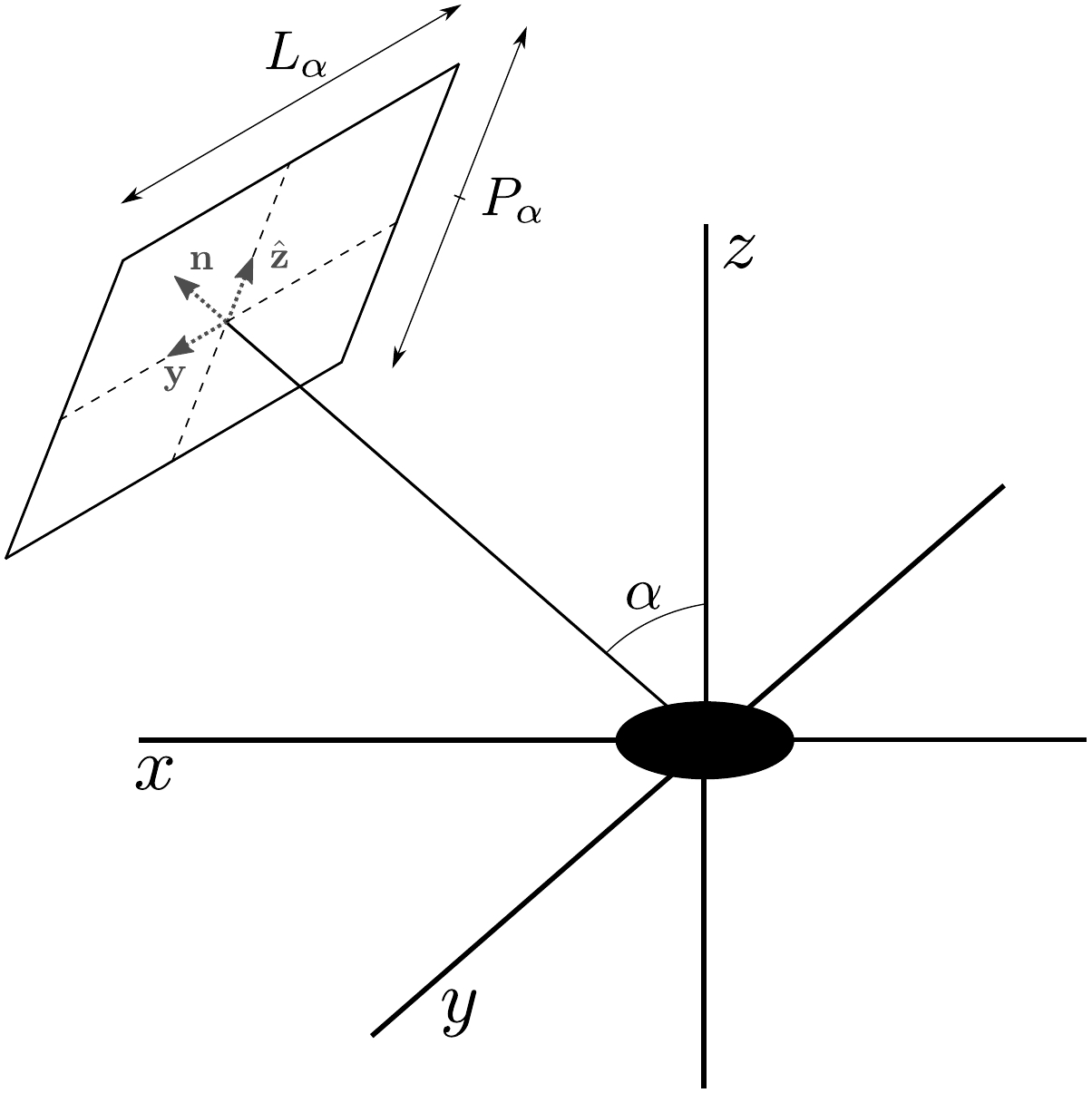}
	\caption{\small Schematic representation of the null plane, spanned by $L_\alpha$ and $P_\alpha$, representing the final conditions of the merger. The black hole rotates along the $z$ axis, and the vector $\mathbf{n}$ denotes the collision direction.}
	\label{fig:LQspace}
\end{figure}

We may label the different generators of the asymptotic null surface by the two conserved quantities $L$ and $Q$ but, as already mentioned, $P_\alpha$ may be a better parameter than $Q$. For the purpose of interpreting $P_\alpha$, and for later visualization of the results, it is convenient to define the following Cartesian-like coordinates
\begin{subequations}
\label{eq:embed}
\begin{align}
x&=\sqrt{r^2+a^2+\frac{2\, m\, a^2 r \sin^2\theta}{\Sigma}}\sin\theta\cos\varphi,\\
y&=\sqrt{r^2+a^2+\frac{2\,m\, a^2 r \sin^2\theta}{\Sigma}}\sin\theta\sin\varphi,\\
z&=r \cos\theta\,.
\end{align}
\end{subequations}
For now we are only interested in their late-time asymptotic limits, obtained by plugging in eqs.~\eqref{eq:IC},
\begin{subequations} 
\begin{align}
x&=\lambda  \sin\alpha- P_\alpha \cos\alpha+\mathcal{O}\left(\frac{1}{\lambda^2}\right)\,,\\
y&=-L \csc\alpha+\mathcal{O}\left(\frac{1}{\lambda}\right)\,,\\
z&=\lambda  \cos\alpha+P_\alpha\sin\alpha+\mathcal{O}\left(\frac{1}{\lambda^2}\right).
\end{align}
\end{subequations}
Let us write this in vector form, $\mathbf{x} = (x,y,z)$ as
\beq
\mathbf{x}=\lambda\, \mathbf{n}+P_\alpha\, \hat{\mathbf{z}}-L_\alpha \, \mathbf{y}\,,
\eeq
where we define
\beq
L_\alpha=L\csc\alpha
\eeq
and we use an orthonormal basis of spatial vectors,
\beqa
\mathbf{n}&=&(\sin\alpha,0,\cos\alpha)\,,\notag\\
\hat{\mathbf{z}}&=&(-\cos\alpha,0,\sin\alpha)\,,\\
\mathbf{y}&=&(0,1,0)\notag\,.
\eeqa
The vector $\mathbf{n}$ points in the asymptotic direction of the null rays, while $\hat{\mathbf{z}}$ and $\mathbf{y}$ span the plane orthogonal to them, \ie\ the spatial sections of the asymptotic null plane. We see that each null ray is assigned rectangular coordinates $(P_\alpha,-L_\alpha)$ on this plane.
Figure~\ref{fig:LQspace} illustrates this construction. 

In general the lines of constant $Q$ are not straight. Instead, since
\beq
P_\alpha^2+\cos^2\alpha\,L_\alpha^2=Q+a^2\cos^2\alpha
\eeq
we see that they are ellipses. If we used $Q$ as a coordinate on the plane, it should run in the range $[-a^2\cos^2\alpha,\infty)$, and we should also take $\sigma=\pm 1$.

There are two particular limits that deserve separate attention:

\paragraph{Aligned merger.}  When $\alpha= 0$ the coordinate $L_\alpha$ degenerates. The $\varphi$-angular momentum $L$ is not a useful parameter since all the geodesics are asymptotically orthogonal to the $\varphi$-rotation plane and thus have $L=0$. The rotation axis of the Kerr black hole is aligned with the collision axis and therefore the $U(1)$ rotational symmetry of Kerr is a symmetry of the whole configuration. In this case the plane is simply covered by using $Q$, or $P_0=\sigma\sqrt{Q+a^2}$, as a radial coordinate, and applying a $U(1)$ rotation to the generators. 

The final conditions for this case can be obtained from expressions~\eqref{eq:IC} by setting first $L=0$ and then $\alpha=0$. The merger in \cite{Emparan:2016ylg}, in which the small black hole was non-rotating, is most simply recovered in this manner. 
\paragraph{Orthogonal merger.} When $\alpha=\pi/2$ we have $P_{\pi/2}=\sigma\sqrt{Q}$ so constant-$Q$ lines are straight in this limit. The merger process does not have any continuous symmetry but possesses a $\mathbb{Z}_2$ reflection symmetry about the equatorial plane $\theta=\pi/2$ ($z=0$).

\subsection{Black hole imaging and shadows}

A very similar analysis of null geodesics is performed when investigating the image of a black hole and the shadow it casts on a background seen by a distant observer. An excellent recent study of this problem can be found in \cite{Pu:2016eml}. 

The family of geodesics considered for that problem is the same as ours, but with a crucial distinction: we cut the surface to the past of caustics and crossovers, since at earlier times the geodesics do not belong to the event horizon. In contrast, for the problem of black hole imaging these points do not play any essential role, and the geodesics are infinitely extended to the past. 

As a consequence there are significant differences in what are the relevant features of the null hypersurface in each case. For instance, the black hole shadow plays no role in our construction, and is instead replaced by a smaller `creaseless disk' that we will discuss below. Also, the surface referred to as the `frozen photon plane' in ref.~\cite{Pu:2016eml} is the same as (a spatial section of) our event horizon, but only until the moment when caustics and crossovers appear.

Given our different motivation and aims, we have preferred to present our construction independently of previous works on black hole imaging such as \cite{Pu:2016eml}. There, following \cite{Chandrasekhar:1985kt}, our parameters $L_\alpha$ and $P_\alpha$ are called $-\alpha$ and $\beta$ and are referred to as the coordinates in the `celestial plane' of the distant observer.  

We cannot help being amused by the fact that an almost (but crucially, not quite) identical construction can be relevant to such different purposes.

\section{Integration of the equations}\label{sec:integresults}

The event horizons are 3-dimensional null hypersurfaces in the 4-dimensional Kerr spacetime. We have constructed different classes of event horizons corresponding to five different collision angles, namely $\alpha=0$, $\pi/8$, $\pi/4$, $3\pi/8$ and $\pi/2$. 

The aligned, axisymmetric configurations $\alpha=0$ are the simplest ones, and although they are rather different than the generic case, they are nevertheless interesting and instructive. 

Orthogonal collisions with $\alpha=\pi/2$ do not have any continuous symmetries. They exhibit all the main qualitative features of the generic case, while the discrete $\mathbb{Z}_2$ reflection symmetry that they possess is useful for their analysis and visualization. 

We have also constructed the event horizon for  collisions with $\pi/8$, $\pi/4$, $3\pi/8$, which are fully generic, but we have found that for these orientations the results are not qualitatively different than for $\alpha=\pi/2$. Therefore we will explain at length the main features of the horizon only for orthogonal collisions, and discuss other intermediate angles in order to see the dependence with $\alpha$ of certain features of the mergers.

The event horizons that we present are generated by integrating numerically backwards in time the equations~\eqref{eq:geo}, subject to the final conditions~\eqref{eq:consts}, which are improved for large but finite final values of $\lambda$ by using \eqref{eq:IC}. The integration can be easily done with the \texttt{NDSolve} function from \textsc{Mathematica}, which uses a fourth-order Runge-Kutta procedure with adaptive step. We stop the integration either when the geodesic is close to the outer Kerr horizon at $r=r_+$, where
\beq
r_+=m+\sqrt{m^2-a^2}\,,
\eeq
or otherwise when it intersects the crease set.\footnote{We stop the integration when $r<r_++10^{-3}$ or when the crossover condition is satisfied. There are generators for which both $r<r_++10^{-3}$ and the crossover condition is close to being satisfied. In these cases we differentiate to what class the generators belong by extrapolating which of the two situations will occur first in the integration (\ie\ at a larger value of $t$).} The crease set consists of \emph{caustics}, which are points where neighboring null rays focus, and of \emph{crossovers}, where two (or possibly more) distinct null rays meet. Generically the crease set on the horizon is a spacelike surface formed of crossover points, and its boundary is a line of caustics \cite{Shapiro:1995rr,Lehner:1998gu,Husa:1999nm}. Therefore, in practice in our numerical construction we can identify the crease set as the set of crossover points where two distinct geodesics meet. In an aligned, axisymmetric collision the crossover points line themselves up along the symmetry axis and they are also caustic points. Let us mention that in the extreme-mass ratio limit the crease set extends infinitely to the past on the horizon.\footnote{Asymptotically it narrows down to a line of caustics of proper length $\sim \sqrt{m}\ln (M/m)$ \cite{Emparan:2016ylg,Hamerly:2010cr}.} 

\begin{figure}[t!]
\centerline{\includegraphics[width=.6\textwidth]{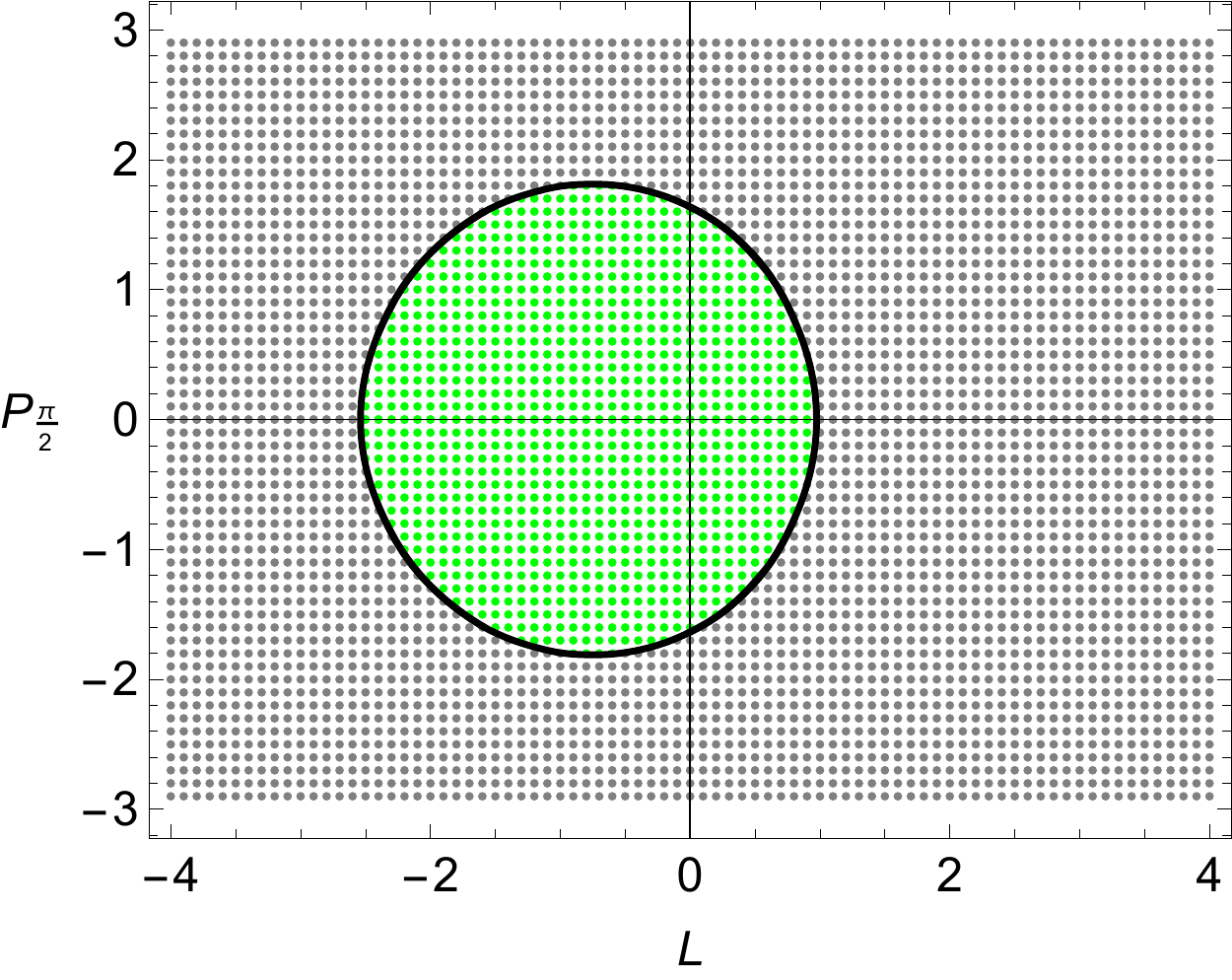}}
\caption{\small Structure of the late-asymptotic null plane $(L_\alpha,P_\alpha)$, for a merger with $a/m=1$ and $\alpha=\pi/2$. Each point corresponds to the position of a null generator of the event horizon at late times. Generators in the green disk do not caustic nor cross (creaseless generators): they originate at asymptotically early times on the Kerr horizon of the small black hole. Gray generators (crease-forming generators) enter the merger event horizon at the crease set at finite time. Those with parameters $(L_{\pi/2},P_{\pi/2})$ and $(L_{\pi/2},-P_{\pi/2})$ cross each other, while rays with $(L_{\pi/2},0)$ caustic. The crease-forming and creaseless classes of generators are separated by the critical black curve $P^c_\alpha(L_\alpha)$. In this specific case the intersections of this curve with the horizontal axis $P_{\frac{\pi}{2}}=0$ are at $L_{\pi/2}=L=L_0^+\approx 0.9$ and at $L=-L_0^-\approx -2.5$. \label{fig:Causticdisk}}
\end{figure}

\paragraph{Creaseless disk.} We find that in the asymptotic plane $(L_\alpha,P_\alpha)$ there is a clear distinction between generators that caustic or cross at finite $\lambda$ (crease-forming generators), and those that never do and instead come from the Kerr horizon at $\lambda\to-\infty$ (creaseless generators). They are separated on the plane by an approximately circular critical curve, $P_{\alpha}^c(L_\alpha)$, which bounds what we call the `creaseless disk' (see fig.~\ref{fig:Causticdisk}). 

Generators with $(L_\alpha,P_\alpha)$ within this disk do not cross nor focus with any other ray but proceed towards $r=r_+$ as $\lambda\to-\infty$, while generators outside the disk always cross others or focus at finite $\lambda$, at a point further from the small black hole the further their parameters $(L_\alpha,P_\alpha)$ are from the disk. 
We can regard fig.~\ref{fig:Causticdisk} as a representation of how the generators of the small black hole horizon fit within the large black hole horizon at late times. In this sense, the disk is the `impression' that the small black hole leaves on the large black hole after falling into it.

Although reminiscent of a `black hole shadow', the creaseles disk is distinct from it. The shadow is cast by those null rays that start from the Kerr black hole and extend to the asymptotic null plane, whether they cross or focus along the way or not. The creaseless disk is composed of the subset of these generators that do not cross nor focus. Therefore it is smaller than the shadow, and contained within it.

The creaseless disk is exactly circular when the system has $U(1)$ axial symmetry, \ie\ when $\alpha=0$, and whenever the rotation vanishes, $a=0$. As the rotation increases the disk gets distorted and displaced off-center, much like the black hole shadow does. The deviations are most marked when $a=m$ and $\alpha=\pi/2$, which is the case shown in fig.~\ref{fig:Causticdisk}. Other creaseless disks will be shown in sec.~\ref{sec:alphapi4}.

The creaseless null geodesics take points that at early times are on the Kerr horizon, with total area $\mathcal{A}_H=8\pi m r_+$, to points in the disk on the plane $(L_\alpha,P_\alpha)$, with total area $\mathcal{A}_\textrm{disk}$. 
The area theorem implies that for every merger we must have
\beq\label{FA}
F_\mathcal{A}\equiv\left.\frac{\mathcal{A}_\textrm{fin}}{\mathcal{A}_\textrm{ini}}\right|_\textrm{creaseless}=\frac{\mathcal{A}_\textrm{disk}}{8\pi m r_+}\geq 1\,.
\eeq
Reference~\cite{Emparan:2016ylg} verified this for the case $a=0$: it found that $F_\mathcal{A}\approx 1.24$. Below we will investigate the influence of rotation on the expansion factor $F_\mathcal{A}$.\footnote{Bear in mind, though, that throughout the merger the total horizon area increases mostly (in fact infinitely in the limit $M\to\infty$ \cite{Emparan:2016ylg}) by the addition of new generators at crease points, rather than by the expansion of those already present. Nevertheless, \eqref{FA} must still hold.}

\paragraph{Visuals.} To visualize the event horizon hypersurfaces we employ two different kinds of plots: 
\begin{itemize}
\item \emph{Projections} of the path traced by the geodesics that belong to the event horizon hypersurface.
\item \emph{Spatial sections} of the event horizon, \ie\ constant-time slices of the merger process.
\end{itemize}

It is important to realize that in our configurations there is a preferred time direction, namely, that of the rest-frame of the small black hole. This is defined by the Killing time $t$ of the Kerr solution, which gives a preferred family of spatial sections. Generically a black hole collision does not possess any timelike isometries, and it is only in the limit $m/M\to 0$ that we have an exact one\footnote{This also implies that gravitational waves are absent, being pushed away from the `Coulomb zone' of the merger that we are focusing on.}. All our plots use this choice of time natural for these configurations.

We show the results in the Cartesian-like coordinates introduced in~\eqref{eq:embed}. These do not give an isometric embedding, but the representation they provide is simple, clear and intuitive. Observe that
\begin{equation}
\sqrt{x^2+y^2}=\sqrt{r^2+a^2+\frac{2\, m\, a^2 r \sin^2\theta}{\Sigma}}\sin\theta
\end{equation}
is the circumferential radius of the $\varphi$ circles. The $z$ coordinate does not have a similar invariant meaning, and measures distances in terms of the radial Boyer-Lindquist coordinate $r$.  Represented in this way, the shape of the Kerr black hole flattens as $a/m$ increases.

In the examples studied below we have varied $0\leq a/m\leq 1$. Generically we find that increasing the rotation only magnifies the differences relative to the non-rotating case $a=0$. In particular, although the effects of rotation in the merger are greatly enhanced in the extremal limit $a/m\to 1$, no qualitative changes appear.

\section{Aligned merger, $\alpha=0$}
\label{sec:alpha0}

Here the collision axis is aligned with the black hole spin. This case (together with the anti-aligned $\alpha=\pi$ case) corresponds to the only rotating configuration for which the collision is axially symmetric. It is therefore a highly non-generic instance of the binary merger, but nevertheless an interesting one to investigate.

Each geodesic is characterized by its impact parameter $P_0$ and angle $\varphi$ in the future asymptotic null plane.
Due to the symmetry of the collision the crossover points are also caustics and the crease set is necessarily a line of caustics located at $\theta=\pi$. 

\subsection{Structure of the event horizon}

\begin{figure}[thbp]
  \subfloat[$t=t_{*} - 7m$\label{subfig:th0-1}]{%
    \includegraphics[width=0.45\textwidth]{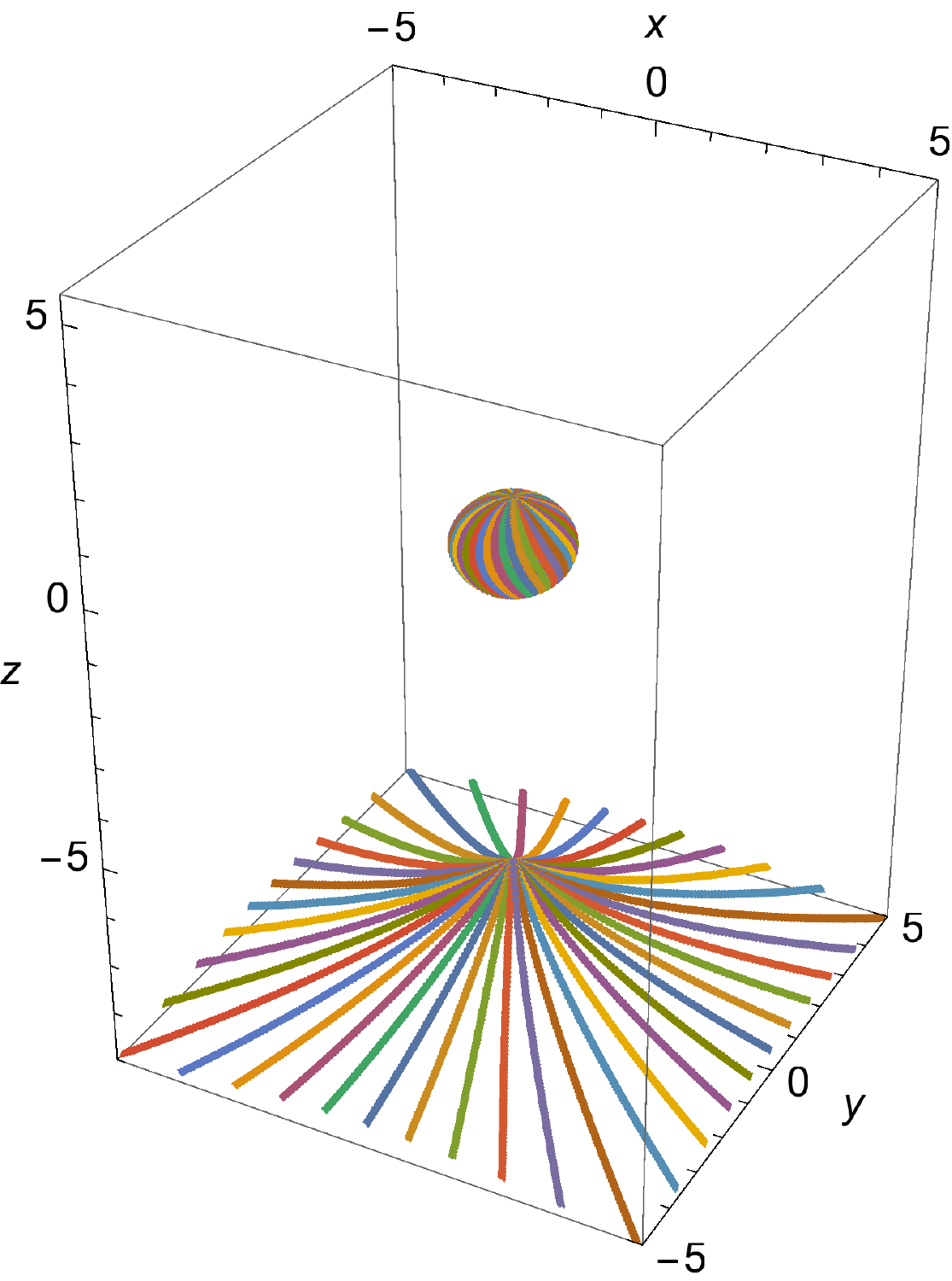}
  }
  \hfill
    \subfloat[$t=t_{*}$\label{subfig:th0-15}]{%
    \includegraphics[width=0.45\textwidth]{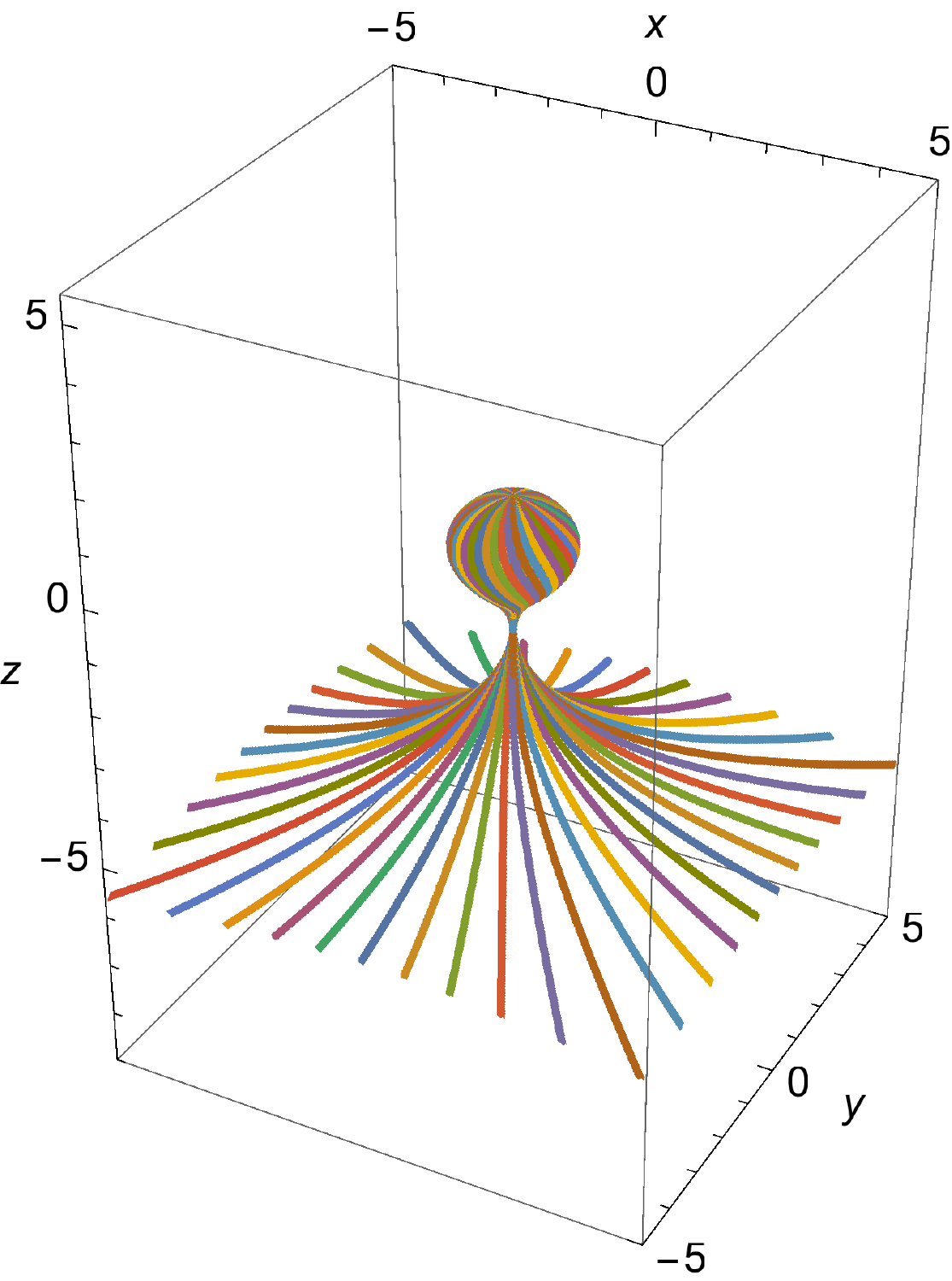}
  } \\
  \subfloat[$t=t_{*} + 6m$\label{subfig:th0-27}]{%
    \includegraphics[width=0.45\textwidth]{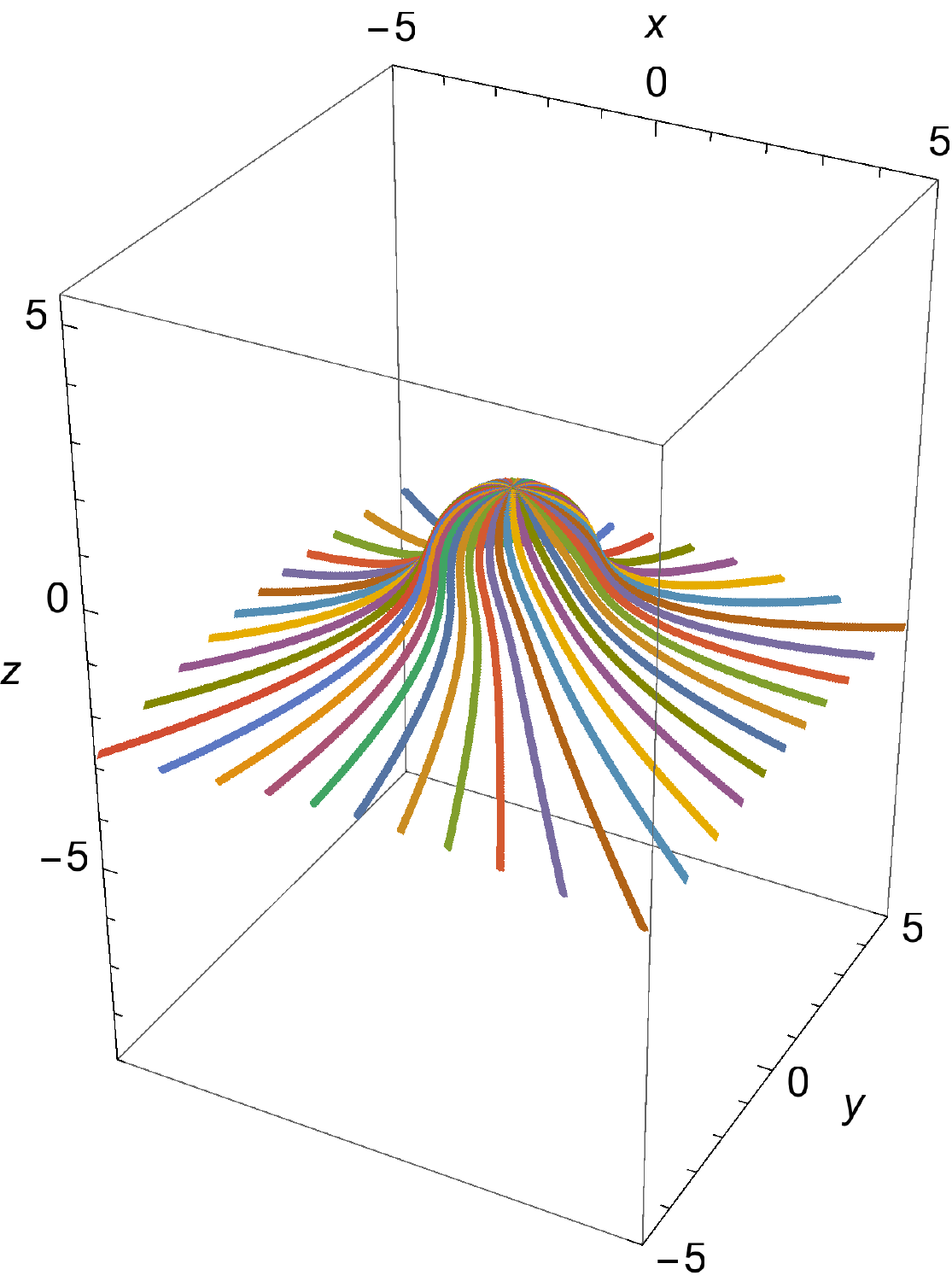}
  }
  \hfill
  \subfloat[$t=t_{*} + 18m$\label{subfig:th0-51}]{%
    \includegraphics[width=0.45\textwidth]{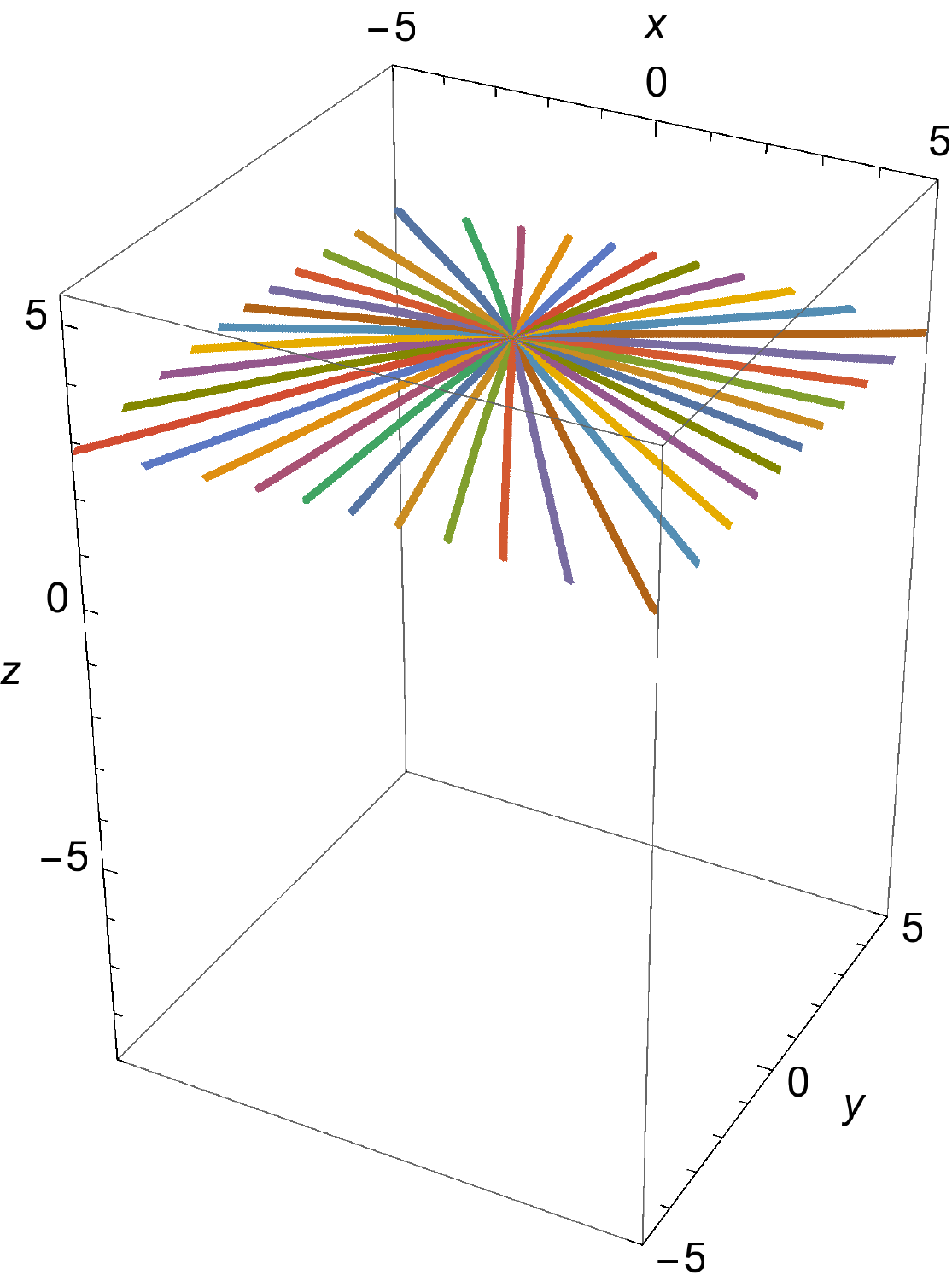}
  }
  \caption[]{\small Constant time slices for the aligned collision, $\alpha=0$,
    case. The time $t=t_{*}$ corresponds to the pinch-on instant at which the two black holes begin to merge. Generators that have the same final azimuthal angle $\varphi$, but different impact parameter $Q$, are shown in the same color. For this case $a/m=0.8$. \label{fig:th0-tslices}}
\end{figure}

Figure~\ref{fig:th0-tslices} shows constant-time slices of this merger for $a/m=0.8$, which is a good representative of an arbitrary rotation. Each line in these plots is a constant-time slice of a null sheet formed of horizon generators that have the same final azimuthal angle $\varphi$ and different impact parameter $P_0$. The picture makes evident that we are in the frame where the small black hole is at rest and the large black hole is approaching it.

At early times (fig.~\ref{subfig:th0-1}) we have two disconnected surfaces: a spheroidal one corresponding to the small Kerr-like black hole, with a horizon distorted by the attraction of the other black hole, and a deformed plane corresponding to the large black hole. The horizons begin to fuse at the time $t=t_{*}$ where they first touch (fig.~\ref{subfig:th0-15}). We will argue in sec.~\ref{sec:locpinch} that this pinch can be studied in a detailed, exact manner. In the subsequent panels, we see how the small black hole is engulfed by the large one, which then relaxes to equilibrium. The rotation of the Kerr black hole is rapidly dissipated by the large black hole, see figs.~\ref{subfig:th0-27} and~\ref{subfig:th0-51}.

Figure \ref{subfig:th0-1} exhibits the feature, present at all times prior to the merger, that both black holes have conical points that correspond to caustic points. In the full spacetime these points form a spacelike line of caustics. This is not generic: in the less symmetric cases with $\alpha \neq 0$ that we study below, the crossovers will form a two-dimensional surface (a very thin strip) instead of a line, and it will wrap around the Kerr black hole. 

\begin{figure}[t!]
\centerline{\includegraphics[width=0.7\textwidth]{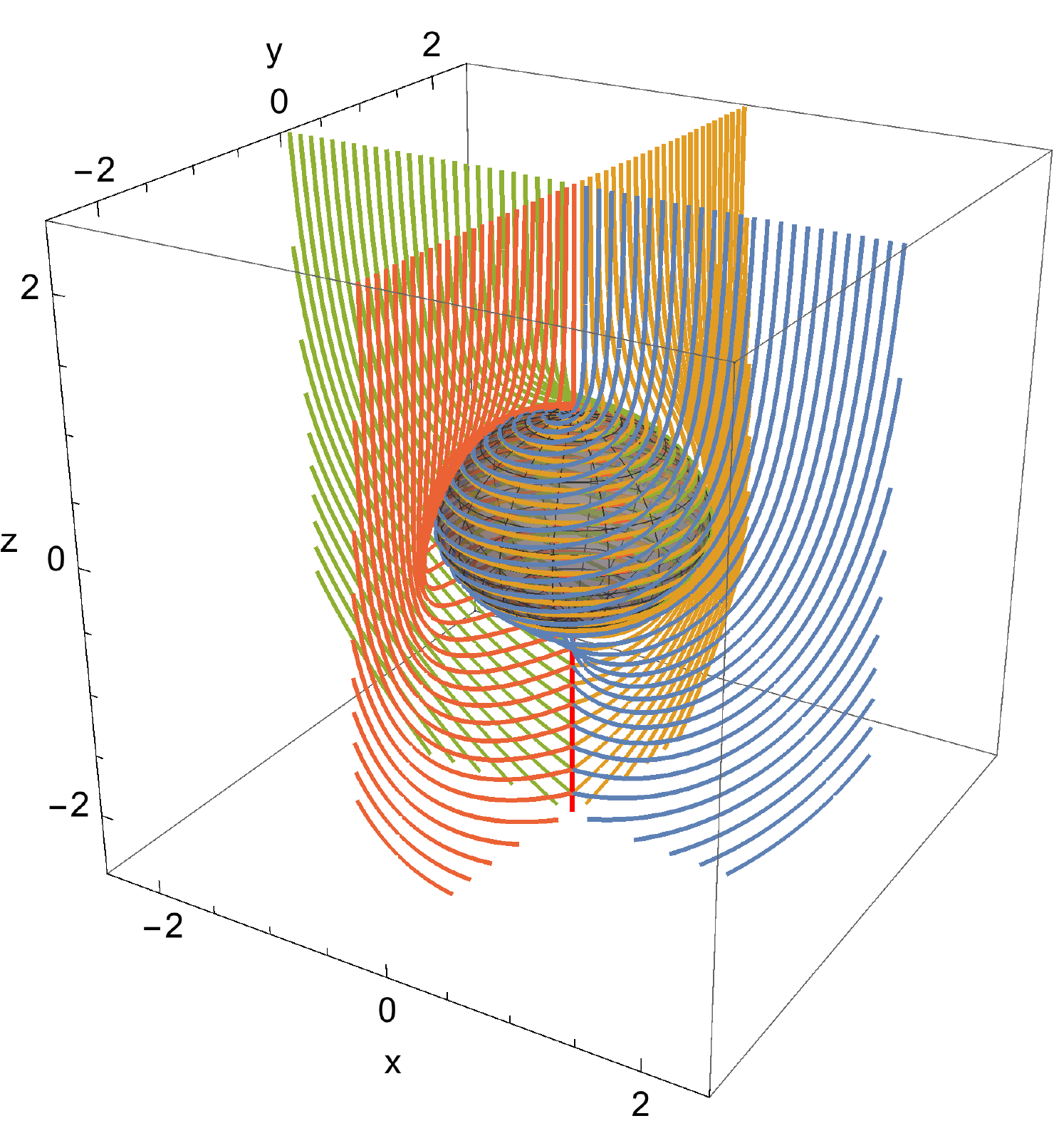}}
\caption[]{\small Generators of the event horizon for the aligned collision projected on the $x,y,z$ space. The red line marks the caustic line; the blue, yellow, green and orange lines are geodesics with different impact parameters and same final azimuthal angles $\varphi$. The gray spheroid marks the Kerr outer horizon. For this case $a/m=0.8$. \label{fig:th0xyz}}
\end{figure}
Figure \ref{fig:th0xyz} shows the projection in the $x,y,z$ coordinates of some horizon generators. Here large values of the $z$ coordinate correspond to late post-fusion times (compare to\ fig.~\ref{subfig:th0-51}). Observe also in fig.~\ref{fig:th0xyz} that generators with sufficiently large impact parameter $P_0$ enter the event horizon at the red caustic line below the Kerr black hole, while generators with smaller impact parameter wrap around the latter, asymptotically corotating with it. That is, on the late-time plane of coordinates $(x,y)$ at large constant $z$, there is a critical circle of radius $P_0^c$ separating caustic and caustic-free (creaseless) generators.

\begin{figure}[thbp]
\centerline{\includegraphics[width=0.5\textwidth]{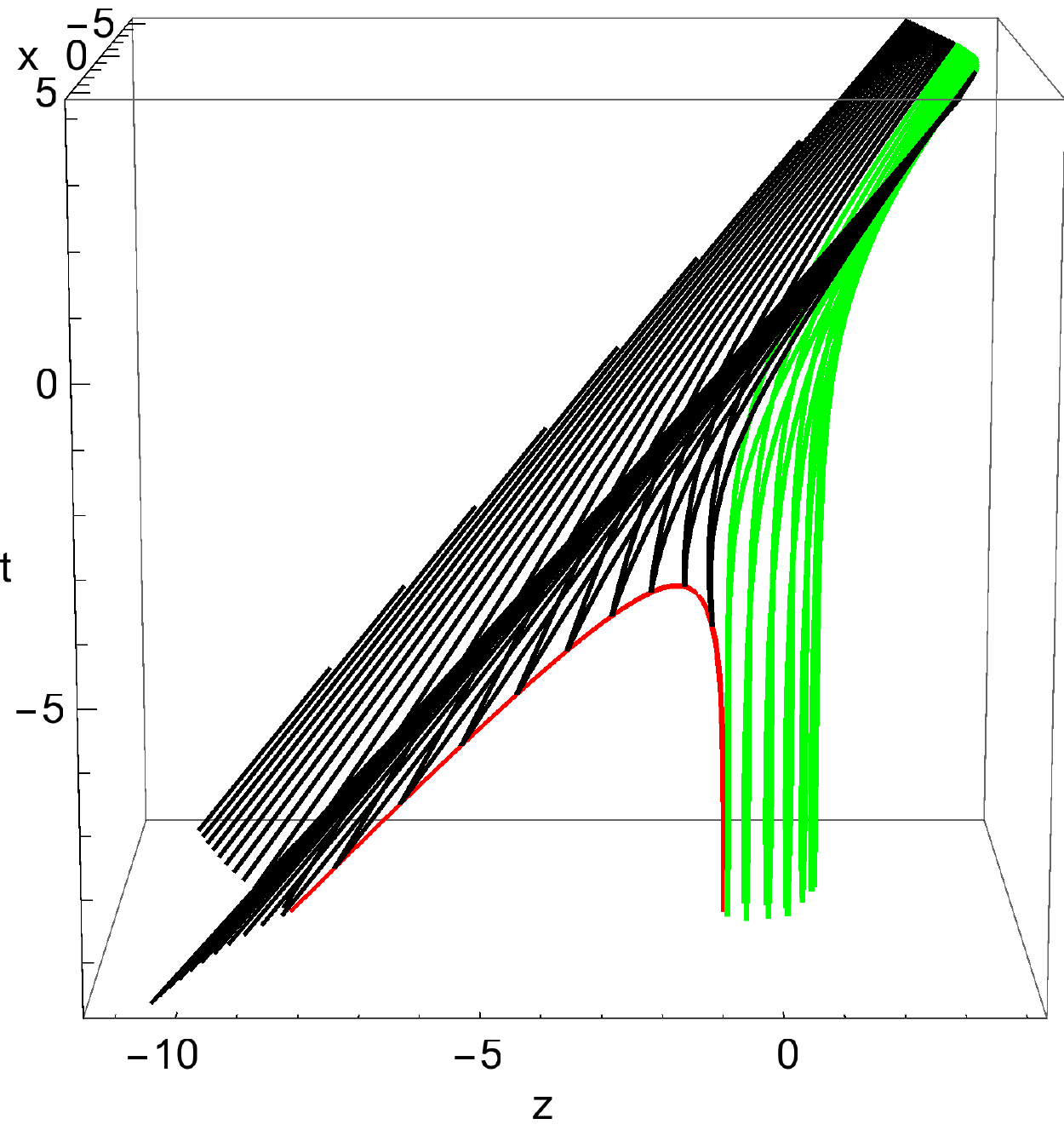}}
\caption[]{\small Generators of the event horizon for the aligned collision projected on the $t,x,z$ spacetime. The red line is the caustic. Black lines are generators that enter the hypersurface through the caustic. Green lines are generators that come from the Kerr horizon. For this case $a/m=0.8$. \label{fig:th0xzt}}
\end{figure}
Figure~\ref{fig:th0xzt} shows a projection of the generators in the $t,x,z$ coordinates, which allows us to view the horizon as a hypersurface in spacetime. Again we distinguish the two different kinds of generators: the black ones enter the event horizon through the caustic line (red), while the green ones do not caustic but come from the Kerr horizon at asymptotic early times. Recall that Kerr generators (null orbits of $\xi = \partial_t + \Omega_H \partial_{\varphi}$) move along parallels of the spheroid with angular velocity $\Omega_H$.

\subsection{Effects of increasing rotation}

Let us now investigate how certain parameters of the merger change as the rotation increases. In order to make a meaningful comparison between black holes with different rotation parameter $a$, we must fix the units, \ie\ we must specify that the black holes we compare have the same size, for some specific notion of size. Two natural choices are to keep the mass $m$ fixed, or alternatively, to keep the horizon area fixed. In the latter case, the unit of length is the area radius 
\beq
r_A=\sqrt{\frac{\mathcal{A}_H}{4\pi}}=\sqrt{2m r_+}\,.
\eeq
It should be clear that as we vary $a$ from $0$ to $m$ the two representations contain the same information. However, they may lend themselves to different interpretations. 
For this purpose it is useful to recall that when $a$ grows with $m$ fixed the Kerr horizon area, and hence its number of generators, decreases. On the other hand, if we keep fixed the horizon area of the small black hole (and hence the number of Kerr horizon generators), then the mass $m$ grows as $a$ increases. 

\paragraph{Expansion of creaseless horizon generators.}
On the future asymptotic plane the generators that do not caustic all lie on a disk of area $\pi (P_0^c)^2$. These null rays originate in the asymptotic past in the Kerr horizon, with area $4\pi r_A^2$. Then the expansion factor \eqref{FA} is 
\beq
F_\mathcal{A}=\lp\frac{P_0^c}{2r_A}\rp^2\,.
\eeq

\begin{figure}[t!]
\centerline{\includegraphics[width=0.5\textwidth]{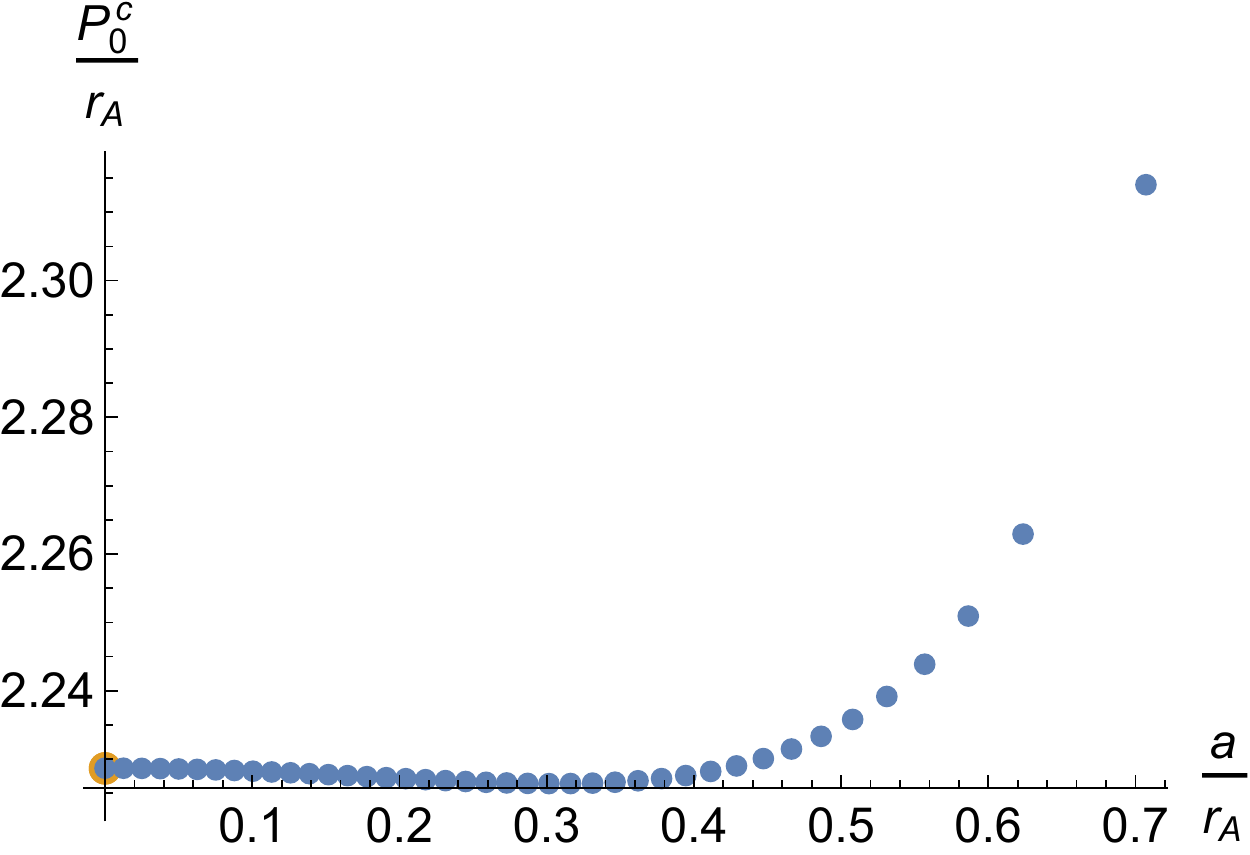}}
\caption[]{\small
Critical impact parameter $P^{c}_{0}$ as a function of rotation for fixed area radius $r_A$ (blue dots). The value $P_0^c=2.23\, r_A$ for $a=0$ was obtained in \cite{Emparan:2016ylg}. Notice that $P_0^c/r_A$  varies by less than $4\%$ over the entire range of rotations. The extremal limit is at $a/r_A=1/\sqrt{2}$.
\label{fig:P0vsa-fixed area}}
\end{figure}

Figure~\ref{fig:P0vsa-fixed area} shows that $P_0^c/r_A>2$ and hence $F_\mathcal{A}>1$ for all values of the rotation, in accord with the area theorem. But less expected is that $P_0^c/r_A$ varies little as $a$ increases. The expansion factor $F_\mathcal{A}$ changes by no more than $8\%$ from the static value $1.24$ even at extremal rotation. That is, the expansion of the creaseless generators is quite independent of the rotation. The slight increase of the expansion at large rotations is presumably due to the larger mass that Kerr black holes of fixed area have for larger $a$: more generators are attracted by the more massive Kerr black hole, so the final disk is larger.\footnote{We do not have any simple explanation for the behavior of $P_0^c/r_A$ at small rotation, but note that a competition between gravitational attraction of mass and centrifugal repulsion of rotation may be at play.} 

\begin{figure}[t!]
\centerline{\includegraphics[width=0.5\textwidth]{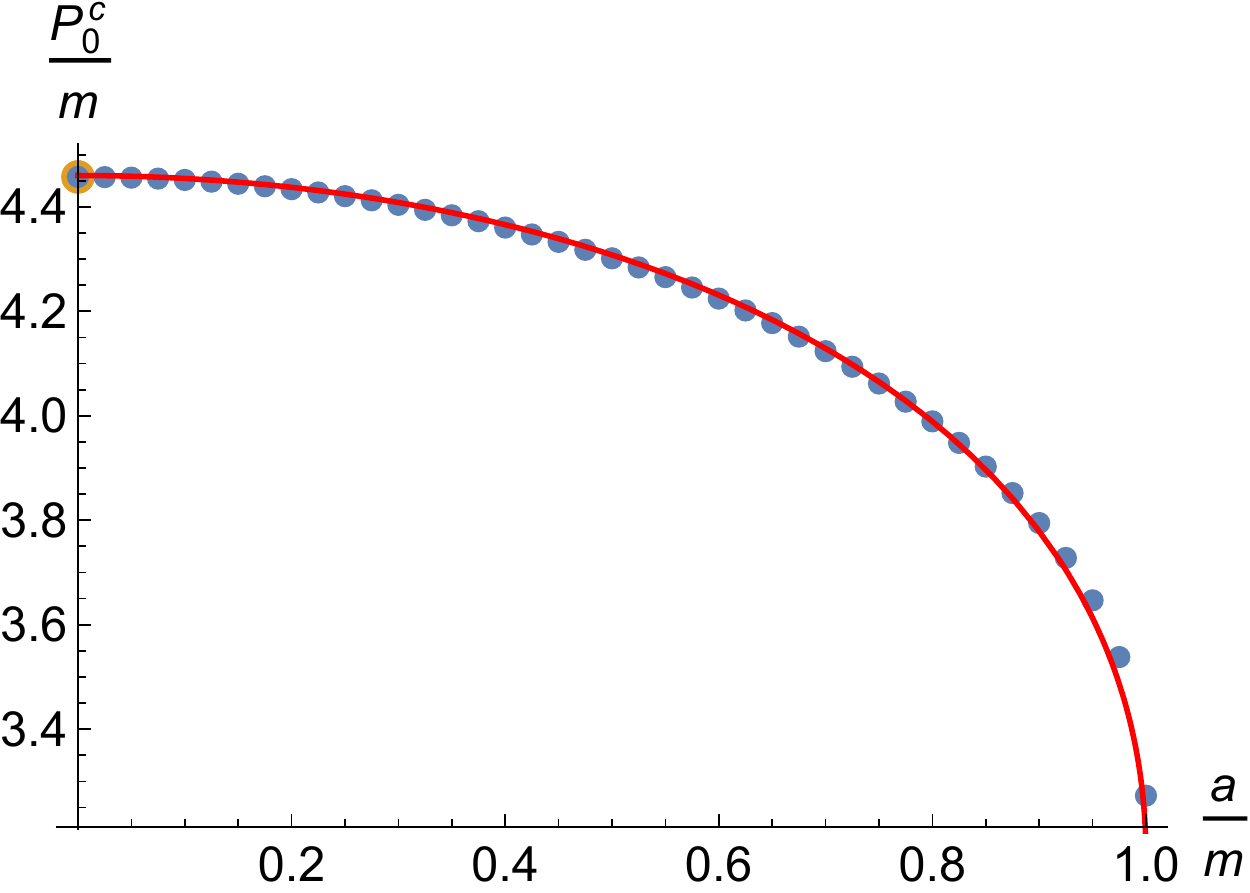}}
\caption[]{\small Critical impact parameter $P^{c}_{0}$ as a function of rotation for fixed mass $m$ (blue dots). The value $P_0^c=4.46\, m$ for $a=0$ was obtained in \cite{Emparan:2016ylg}.  The red line shows the Kerr area-radius $r_A/m$ as a function of $a/m$, multiplied by $2.23$ so the two curves agree at $a=0$. We see that the variation of $P_0^c$ with the rotation is well predicted by the area radius. This implies that creaseless generators expand by a factor that is quite independent of the rotation. \label{fig:th0P0cvsa-fixedmass}}
\end{figure}

The approximate constancy of $P_0^c/r_A$ with rotation, as observed in fig.~\ref{fig:P0vsa-fixed area}, implies that when we compare rotating black holes of the same mass, the values of the critical impact parameter $P^c_{0}$ must be well predicted by their respective Kerr horizon-area radii (up to an overall factor fixed at a reference rotation, \eg\ $a=0$). This expectation does indeed bear out, as shown in fig.~\ref{fig:th0P0cvsa-fixedmass}. Even at the maximum rotation, where we have $P_0^c=3.27\, m$, the red curve yields the value $4.46\, m/\sqrt{2}=3.15\, m$.

\paragraph{Relaxation time.} Let us now study how the rotation of the small black hole affects the time that the large black hole takes to relax back to equilibrium after the fusion. It turns out that, even if the relaxation time of any particular merger is not defined unambiguously, it is nevertheless possible to give a meaningful notion of the difference between the relaxation times of two mergers with different values of $a/m$.

To see how this is possible, recall first that in these mergers there is a preferred choice of time, namely the Killing time $t$ of the Kerr solution. For each merger we can compute the time $t_*$ at which the two black holes first touch---the pinch-on instant. The event horizon afterwards relaxes towards uniform equilibrium. The specific value of $t_*$ is arbitrary: we could change it by choosing a different value of the integration constant $t_\infty$ in \eqref{eq:tinf}. However, since we set the value $t_\infty=0$  independently of $a/m$, then the difference in the values of $t_*$ for two mergers with different $a/m$ is meaningful: it provides a sensible characterization of the difference in the time that each configuration takes to relax to the final equilibrium. 

We take the non-rotating $a=0$ case as a reference. Then, if the difference
\beq
t_*(0)-t_*(a)
\eeq
is positive (negative) we conclude that the merger with rotation $a$ takes a longer (shorter) time to relax than the one for Schwarzschild. 

We make the comparison in the two manners discussed above: fixing the mass or the horizon area.\footnote{Observe that if we compare the absorption of two black holes of different sizes, \eg\ non-rotating ones with different masses, then by scaling symmetry the ratio of their relaxation times must be the same as the ratio of their masses.} The relative relaxation time increases or decreases with $a$ depending on this choice.

\begin{figure}[thbp]
  \subfloat[fixed area\label{fig:th0Dtvsa-fixedarea}]{%
    \includegraphics[width=0.47\textwidth]{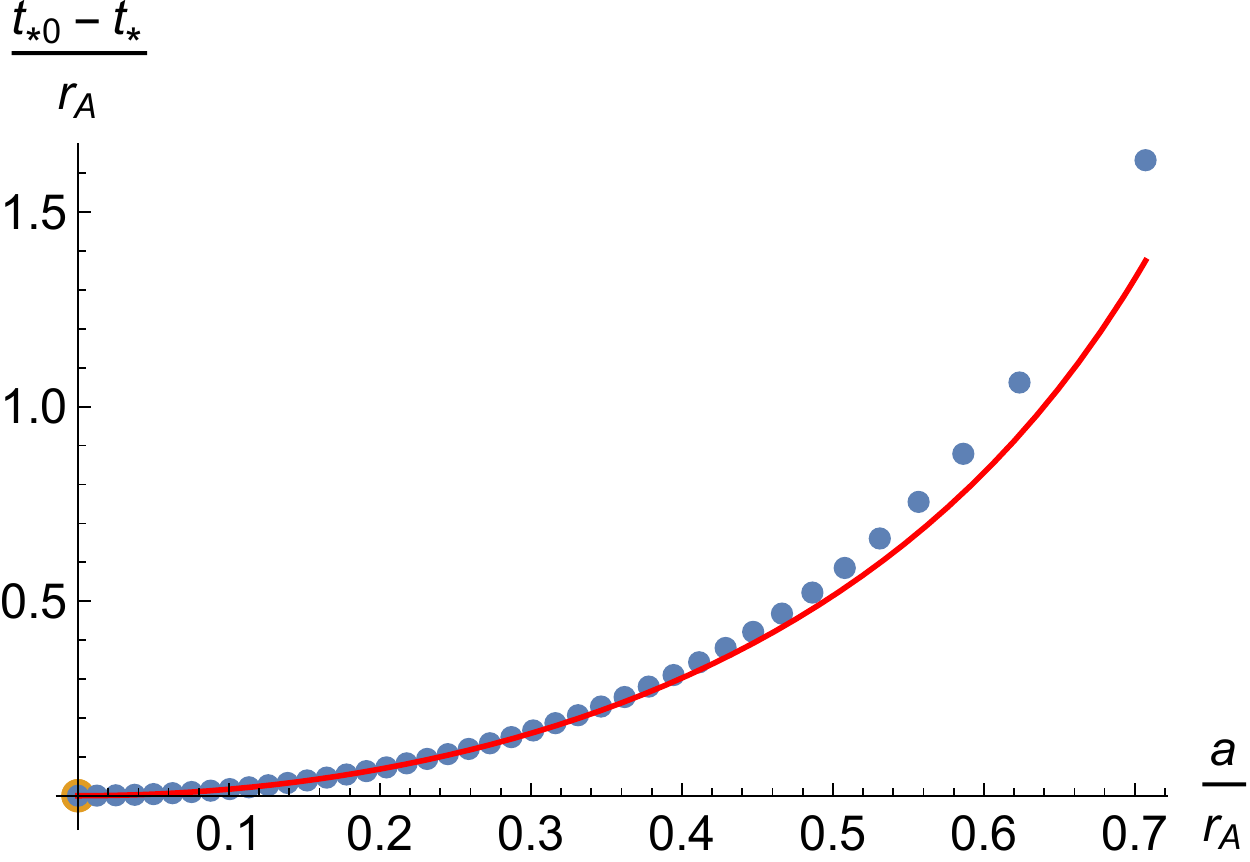}
  }
  \hfill
  \subfloat[fixed mass\label{fig:th0Dtvsa-fixedmass}]{%
    \includegraphics[width=0.47\textwidth]{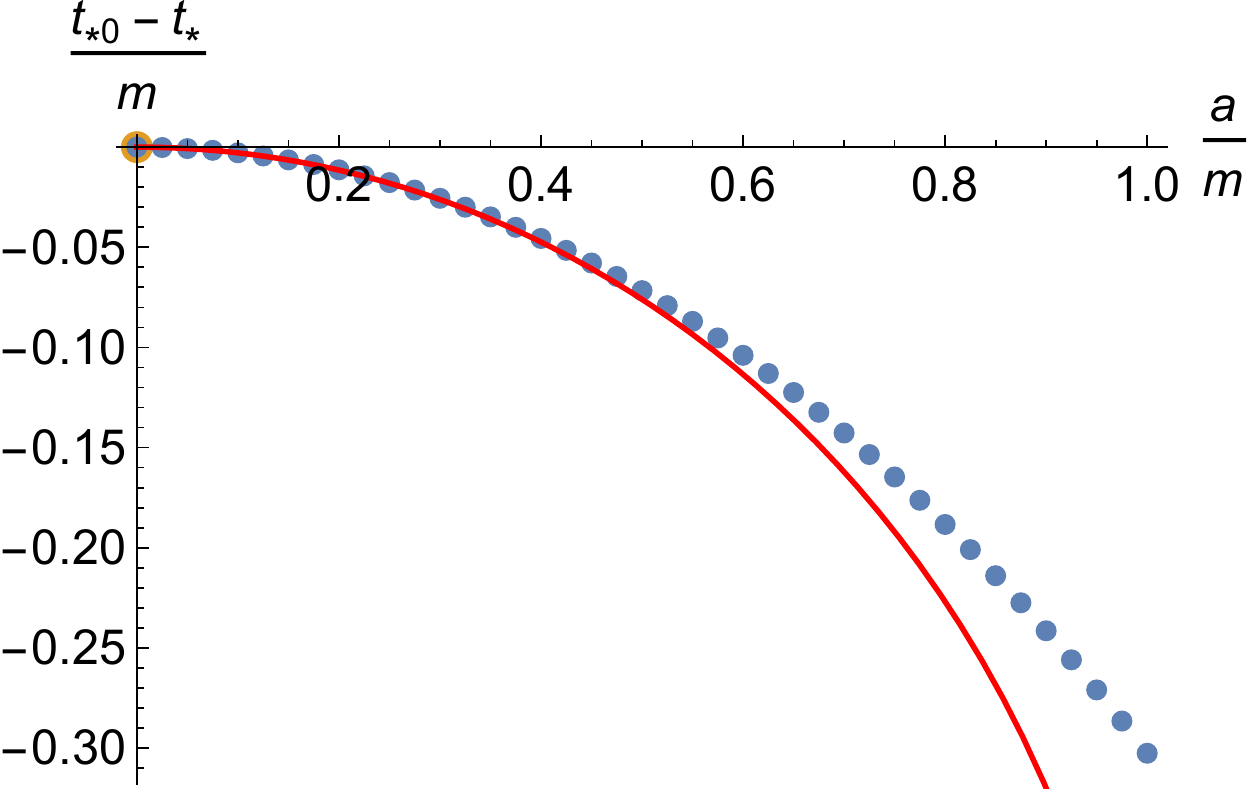}
  }
  \caption{\small Relative relaxation times (with respect to Schwarzschild) as a function of the small black hole rotation $a$. In fig.~(a) we compare black holes with the same horizon area. The red curve shows the variation in the mass of the black holes, multiplied by a factor to match the second derivative of the relaxation time at $a=0$ (other choices of this factor that give a better fit at large rotation are also possible). The good fit means that the growth  of the mass with $a$ at fixed area predicts very well the variation in relaxation times. 
In fig.~(b) we compare black holes with the same mass $m$. We see that the variation in relaxation times can sensibly be attributed to the decrease of the initial Kerr horizon area with $a$ at fixed $m$ (red line, similarly fitted by a single parameter at small rotation). That is, it is due to the reduction in the number of generators that the large black hole has to absorb from the small black hole.}\label{fig:th0Dt*}
\end{figure}

Figure~\ref{fig:th0Dtvsa-fixedarea} shows that as the rotation increases keeping the area fixed, the merger takes longer to relax. This can be understood as the effect that the large black hole takes a longer time to absorb and digest an increasing mass as $a$ grows for fixed area. The red line shows that this interpretation is quantitatively correct: the increment in mass for fixed horizon area predicts very well the variation in relaxation time.\footnote{In our numerical calculations, which involve a certain inaccuracy from the finiteness of the initial value of $\lambda$, we have been careful to check that this numerical imprecision is much smaller than the values of $t_*(0)-t_*(a)$ that we find: the first $a$-dependent term is $\mathcal{O}(\lambda^{-2})\sim 10^{-4}/m$ for $\lambda_\text{initial}=100$.}

In contrast, fig.~\ref{fig:th0Dtvsa-fixedmass} shows that when we increase the rotation keeping the mass fixed, the merger occurs faster. This is expected: the large horizon has to accommodate fewer generators that come from the small black hole, and can do it more quickly. The red line for the area variation in the figure shows that this interpretation is quite accurate.

\section{Orthogonal merger, $\alpha=\frac{\pi}{2}$}\label{sec:alphapi2}

In this case the collision axis is perpendicular to the rotation axis of the Kerr black hole. The reflection symmetry about the equatorial plane in this configuration facilitates the localization of the crease set. Still, the structure of the latter remains qualitatively similar to (and more easily visualizable than) that of the generic case $0<\alpha< \pi/2$.

The null geodesics that rule the horizon hypersurface are labeled by $L$ and $P_{\pi/2}$. For clarity in this section we omit the subindex ${}_{\pi/2}$ in $P$.

\subsection{Constant-time slices and toroidal topology}\label{subsec:constt}

\begin{figure}[tbp]
  \subfloat[$t=t_{*}-8m$\label{subfig:thpi2-1}]{%
    \includegraphics[width=0.45\textwidth]{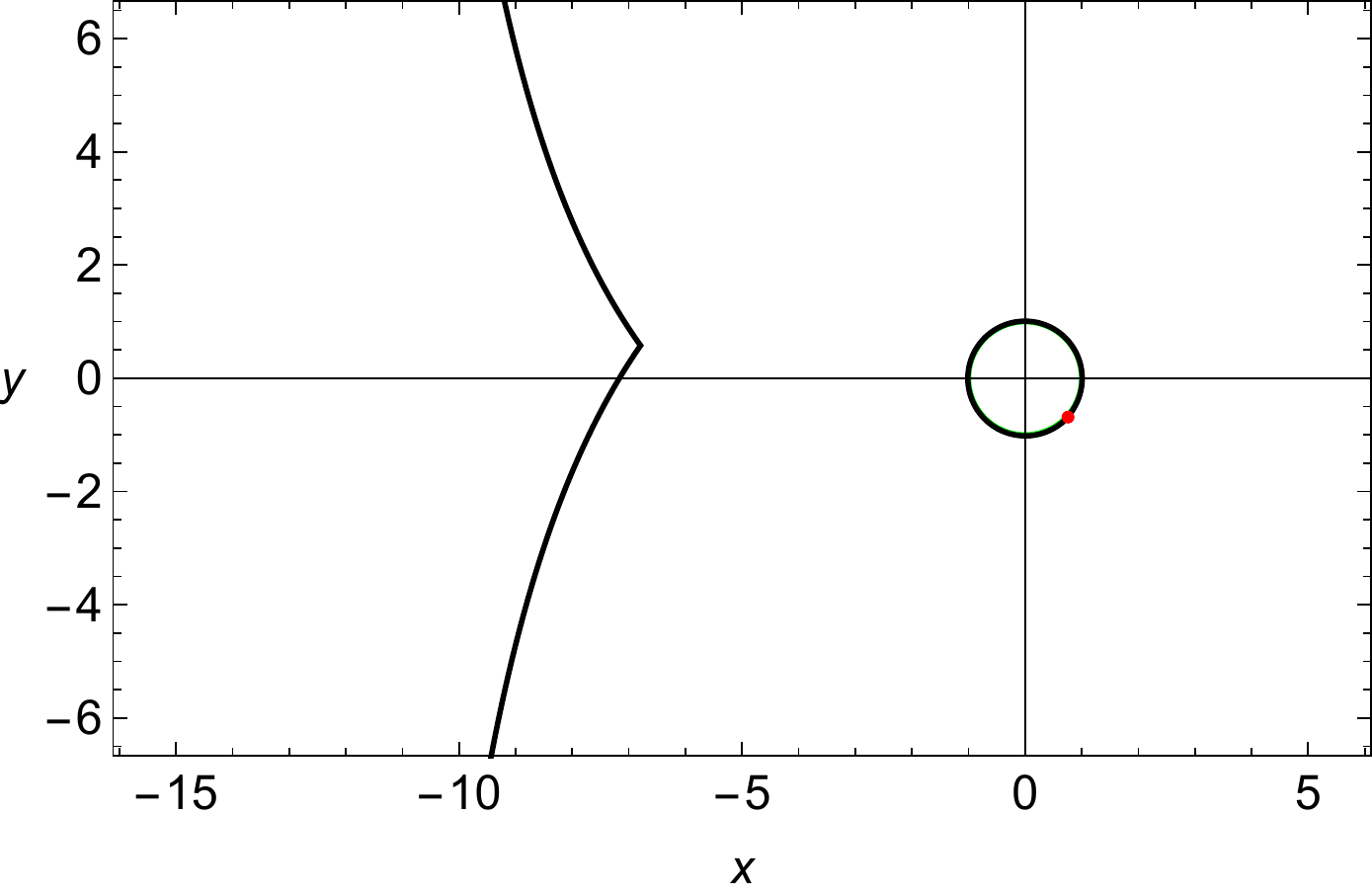}
  }
  \hfill
    \subfloat[$t=t_{*}-4m$\label{subfig:thpi2-2}]{%
    \includegraphics[width=0.45\textwidth]{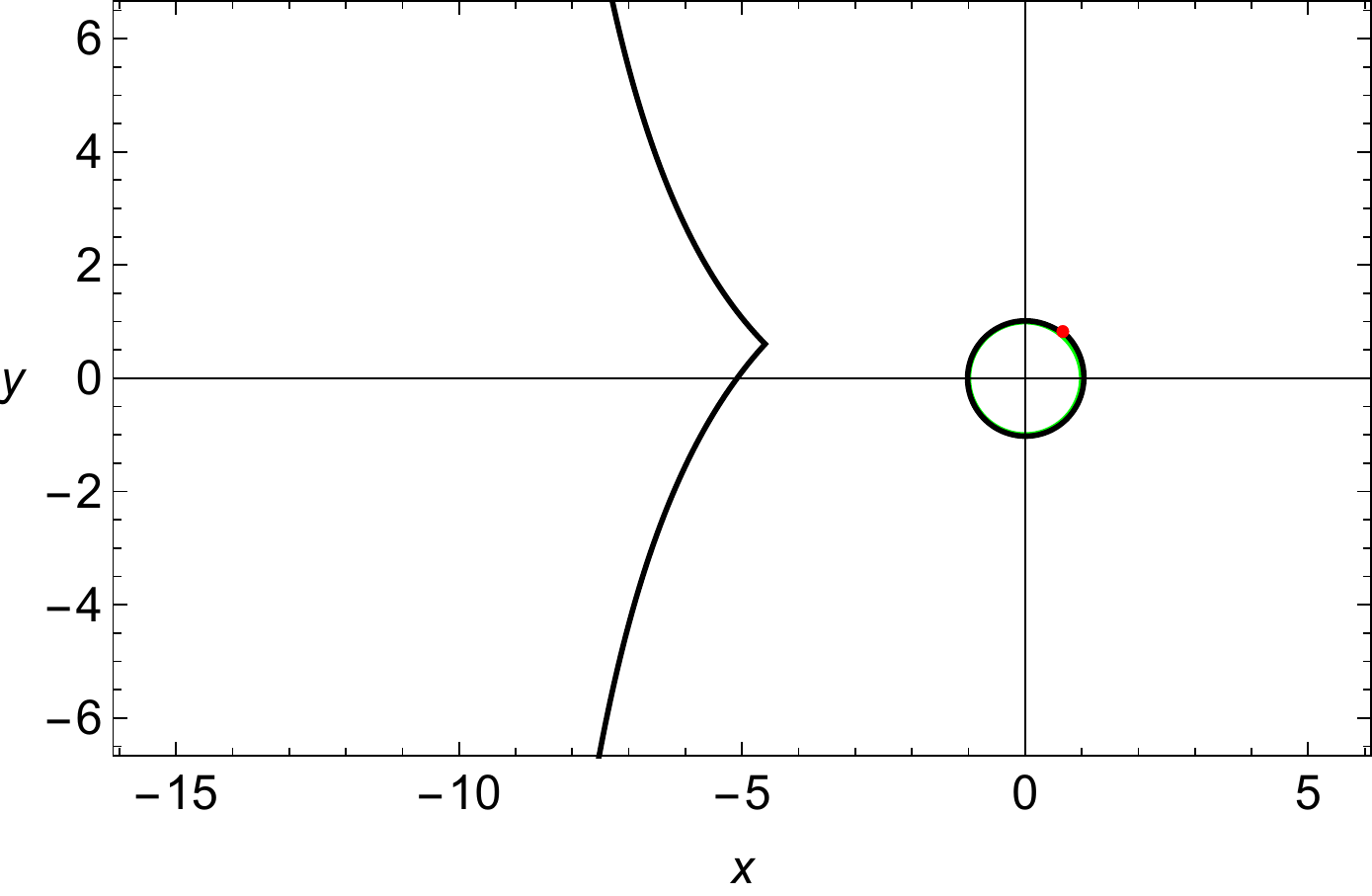}
  } \\
  \subfloat[$t=t_{*}$\label{subfig:thpi2-3}]{%
    \includegraphics[width=0.45\textwidth]{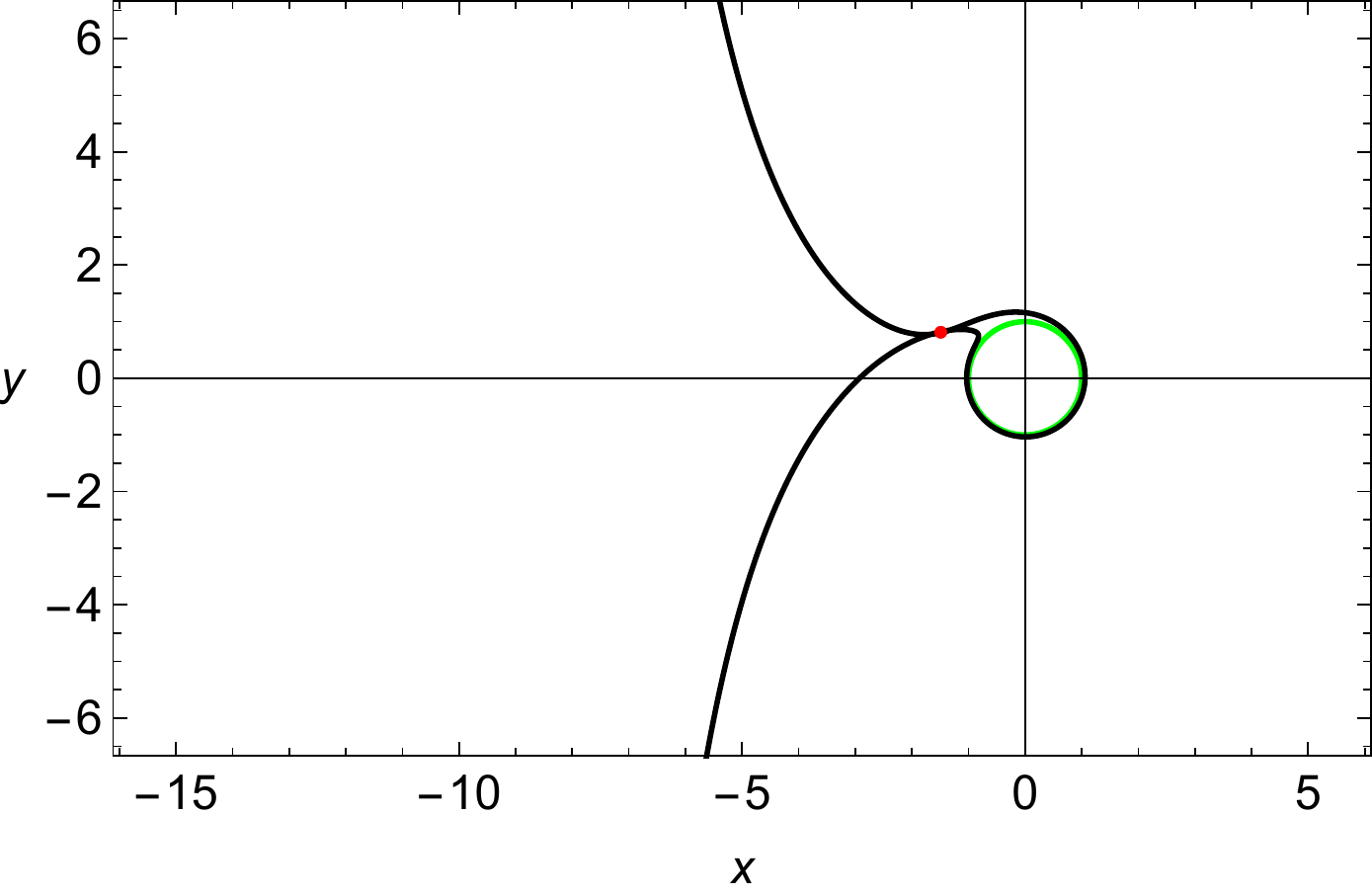}
  }
  \hfill
  \subfloat[$t=t_{*}+2m$\label{subfig:thpi2-4}]{%
    \includegraphics[width=0.45\textwidth]{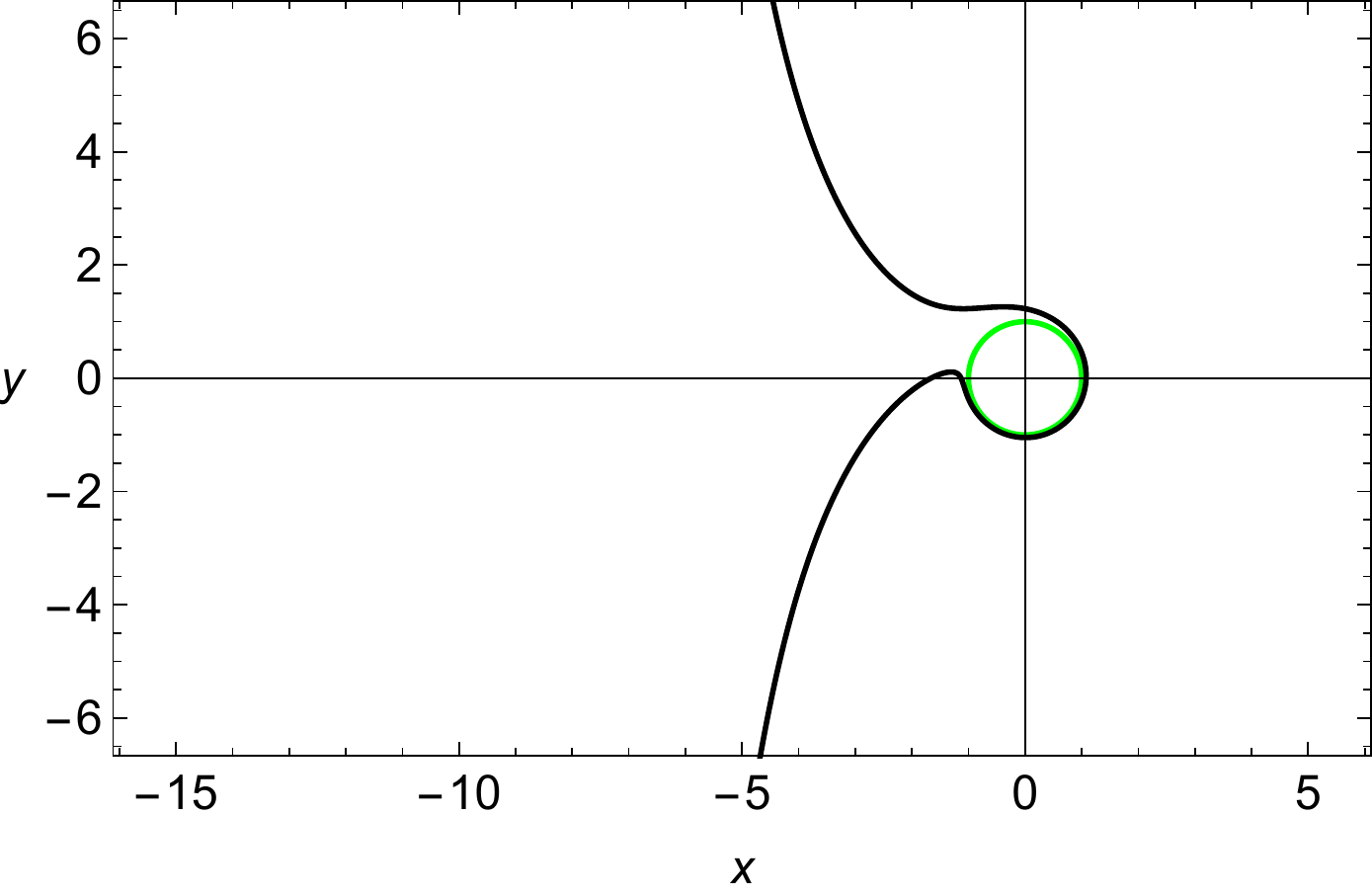}
  }
  \caption[]{\small Equatorial constant-time slices for the orthogonal collision, $\alpha=\pi/2$. The time $t=t_{*}$ corresponds to the pinch-on instant at which the two black holes first touch. Black lines depict the event horizon of the merger, whereas the green circle shows the conventional Kerr black hole horizon. The red dot is the position of the crease (almost a caustic cone) in the small black hole. Here $a=m$ and the small black hole rotation is counter-clockwise. There is structure in the event horizon near the conical tips that is not resolved in these pictures (see next figure). \label{fig:thpi2-tslices}}
\end{figure}

Before we go on to discuss the relatively complex crease set structure, let us present the more readily grasped constant-time slices of the merger.  Figure~\ref{fig:thpi2-tslices} shows the time evolution of the event horizon on equatorial plane sections. 

In a constant-time slice, unless we use high enough resolution (on the order of a percent of $m$), the creases look like caustic points: they are the conical points in the large black hole horizon, and the red dots (also conical) in the small black hole. 

These cones, in both the large and small black holes, move in time due to the rotation, but they do so in different ways. In fig.~\ref{fig:thpi2-tslices} we see that the red dot tends to corotate (counter-clockwise) with the small black hole: this is the expected dragging. But in the large black hole what we see is the teleological nature of the event horizon. The large black hole is not dragged by the rotation of the small one as ordinary matter would do: it is in fact anti-dragged. Since the event horizon is defined at future infinity, the large black hole ``knows'' that it has to relax into a planar horizon and anticipates the encounter with the small black hole by moving clockwise towards the $y>0$ region to meet it.

\begin{figure}[t!]
\centerline{\includegraphics[width=0.8\textwidth]{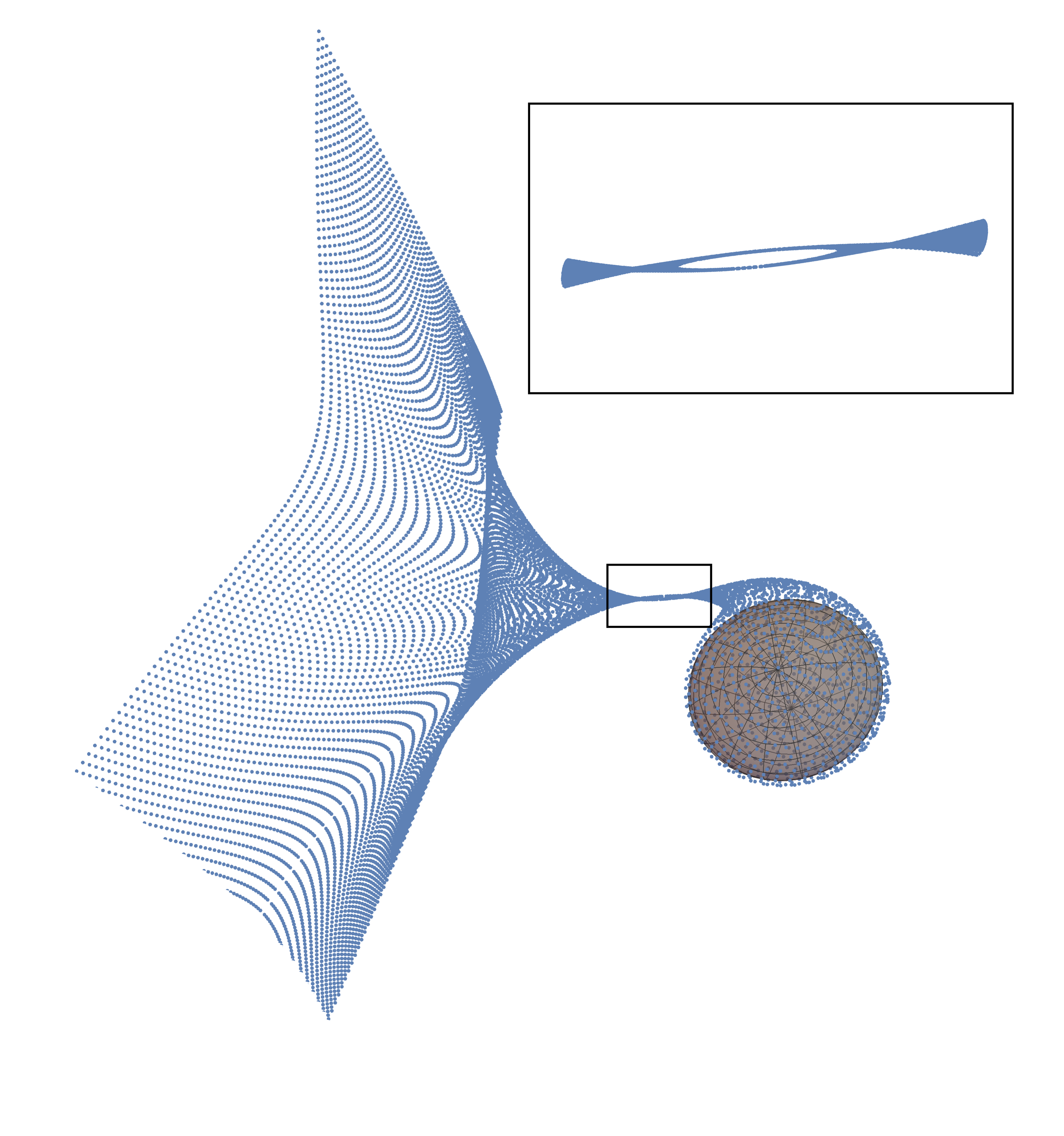}}
\caption[]{\small Constant time slice of the event horizon at a time when the two black holes are fusing. The inset shows a zoom-in of the thin (toroidal) region. In gray we depict the Kerr horizon. Each dot corresponds to the position of a generator at this given time. For this case, with $a=m$, the maximum size of the hole is about $0.1\, m$ in length and $0.01\, m$ in width (in the preferred time-slicing of this system).\label{fig:torus}}
\end{figure}
If we increase the resolution we can see that, before the merger, the horizon near each of the conic points is resolved into chiseled shapes, each one a cone of which the tip is not a point but a very short (and curved) segment. These two segments are curved like $\subset\,\,\supset$, and when they meet and fuse they give rise to a circle: a hole on the horizon. This is visible in fig.~\ref{fig:torus}. If we compactify the large black hole horizon by adding the points at infinity (so that at early and late times its topology is spherical), we can say that these sections of the merger horizon have the topology of a torus.

The occurrence of toroidal topology in the spatial sections of a generic, non-axisymmetric event horizon during the instants close to merger has been observed and discussed in  \cite{Hughes:1994ea,Shapiro:1995rr,Husa:1999nm,Siino:2004xe,Siino:1997ix,Siino:2005nq,Ponce:2010fq,Cohen:2011cf,Bohn:2016soe}. These toroidal spatial sections of the event horizon are actually a gauge-dependent feature, since one can always choose different spacelike slicings where the topology remains spherical. The invariant, gauge-independent statement is that the crease set is a two-dimensional spatial surface instead of a spatial line. It is this that allows spacelike slicings with toroidal sections of the event horizon \cite{Shapiro:1995rr,Lehner:1998gu,Husa:1999nm}. In the configurations we study, the toroidal topology is readily visible in the preferred, Killing time slicing.

In order to better understand this phenomenon in our construction, we need to turn to a discussion of its crease set.

\subsection{Crease set and horizon structure}

Given the $\mathbb{Z}_2$ symmetry of the configuration, the crossovers must occur in the equatorial plane $\theta=\pi/2$, $z=0$, between pairs of generators with parameter values $(L,\pm P$). To find their locus we integrate generators that at late times lie away from the equatorial plane, \ie have $P\neq 0$, and then stop the integration once they reach the equatorial plane. 

\begin{figure}[t!]
\centerline{\includegraphics[width=0.8\textwidth]{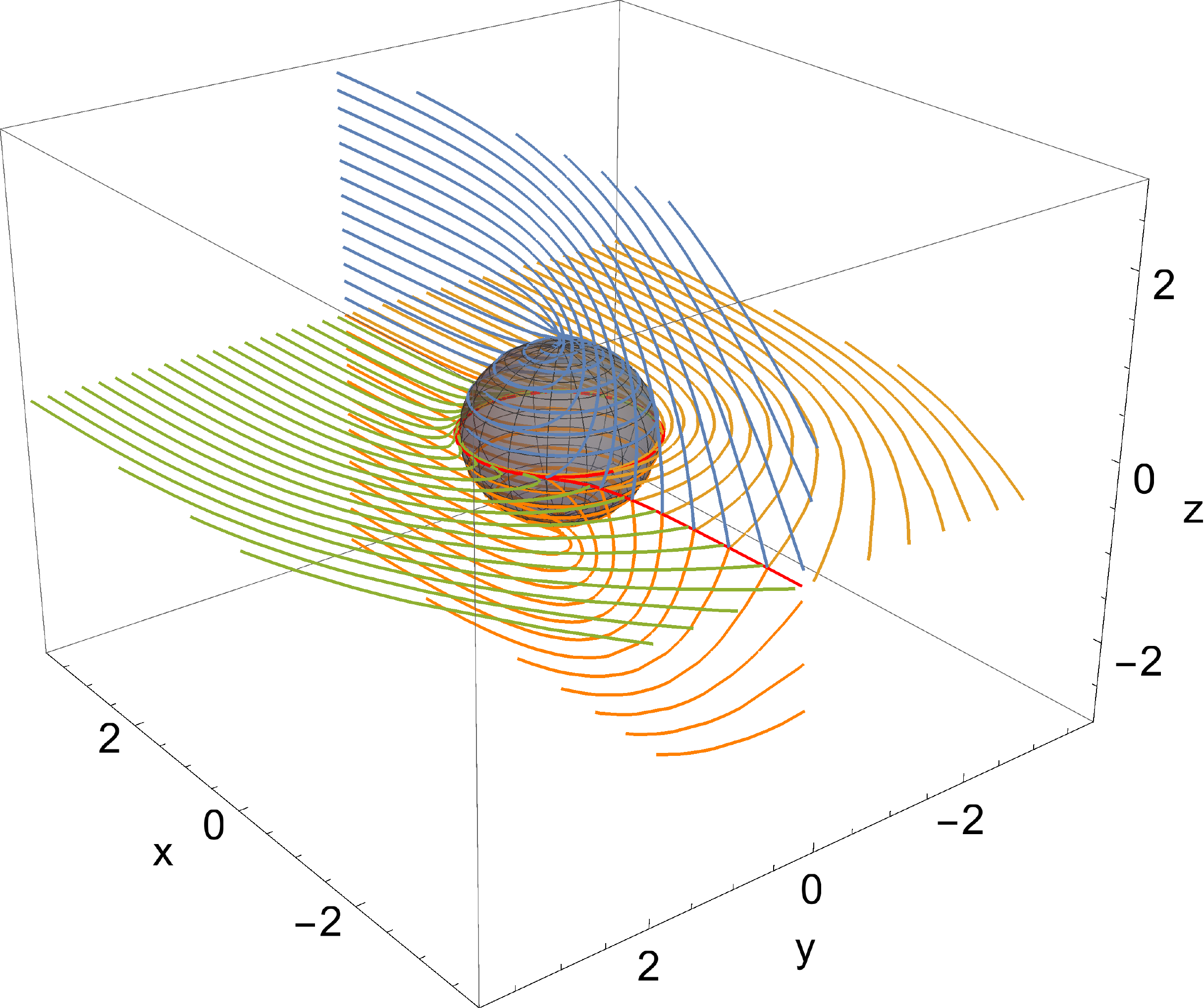}}
\caption[]{\small Generators of the event horizon for the orthogonal collision, projected on the $x,y,z$ space. The impact parameters of the generators at $x\to\infty$ are $P$ in the $z$ direction, and $L$ in the $y$ direction. The red line marks the crease set strip, which winds around the Kerr horizon; its thickness on the $z=0$ plane, not visible in the plot, decreases to zero as $x\to-\infty$ and as the Kerr horizon is approached. The blue and orange lines have $L=0$, whereas green and yellow lines have $P=0$. The gray spheroid marks the Kerr event horizon. For this case $a/m=1/2$. \label{fig:thpi2xyz}}
\end{figure}

In figure~\ref{fig:thpi2xyz} we show, for $a/m=1/2$, generators with $L=0$ (blue and orange lines), which cross whenever $|P|>P_c(0)$, and generators with $P=0$ (green and yellow lines), which cross for positive values $L>L_{0}^+$ and negative values $L<-L_0^-$. In our (as viewed from above) counter-clockwise rotating horizon, we have $L_0^->L_0^+$ (see also fig.~\ref{fig:Causticdisk}, although that one is for $a/m=1$).

\begin{figure}[t!]
\centerline{\includegraphics[width=0.7\textwidth]{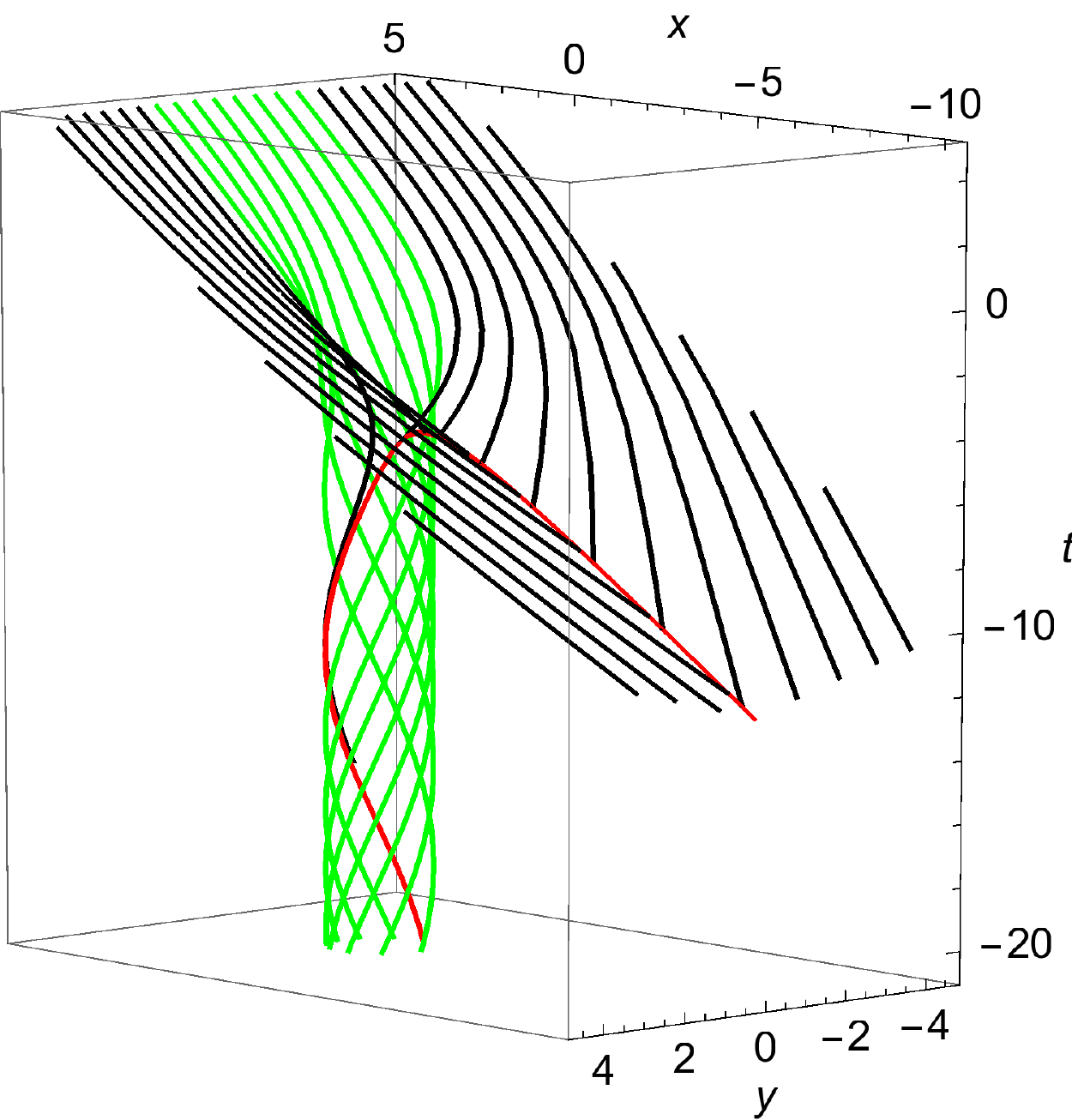}}
\caption[]{\small Generators of the event horizon for the orthogonal collision, projected on the $t,x,y$ spacetime. The red line marks the crease set strip. Its thickness in the $y$ direction, not visible in this plot (but zoomed up in fig.~\ref{fig:causticsurftxy}), decreases to zero towards the asymptotic past. Black lines correspond to generators that enter the hypersurface through the caustic, whereas green lines are generators that come from the Kerr horizon. For this case $a/m=0.5$. \label{fig:thpi2xyt}}
\end{figure}

The crossover points lie on the equatorial plane, where they form not a line but a two-dimensional surface in the shape of a strip (at its boundaries, the generators with $P=0$ and $L>L_{0}^+$ or $L<-L_0^-$ form two caustic lines). This crease set strip is very narrow (width not more than a few percent of $m$), and it shrinks down to zero width towards its two asymptotic endpoints. One of these directions is away from the small black hole towards $x\to -\infty$, where it approaches the line $y=0$. The other direction is towards the small black hole. In this direction the strip winds an infinite number of times around the Kerr horizon, asymptotically corotating with it as its width shrinks to zero; see fig.~\ref{fig:thpi2xyt}. 

Drawing geodesics in this manner, we identify on the asymptotic plane $(L,P)$ the critical curve $P_c(L)$ (black line in fig.~\ref{fig:Causticdisk}) that separates the crease-forming generators that cross on $z=0$ (gray), from the creaseless ones that never cross and asymptote to the Kerr horizon at $t\to-\infty$ (green).

\begin{figure}[t!]
\centerline{\includegraphics[width=.9\textwidth]{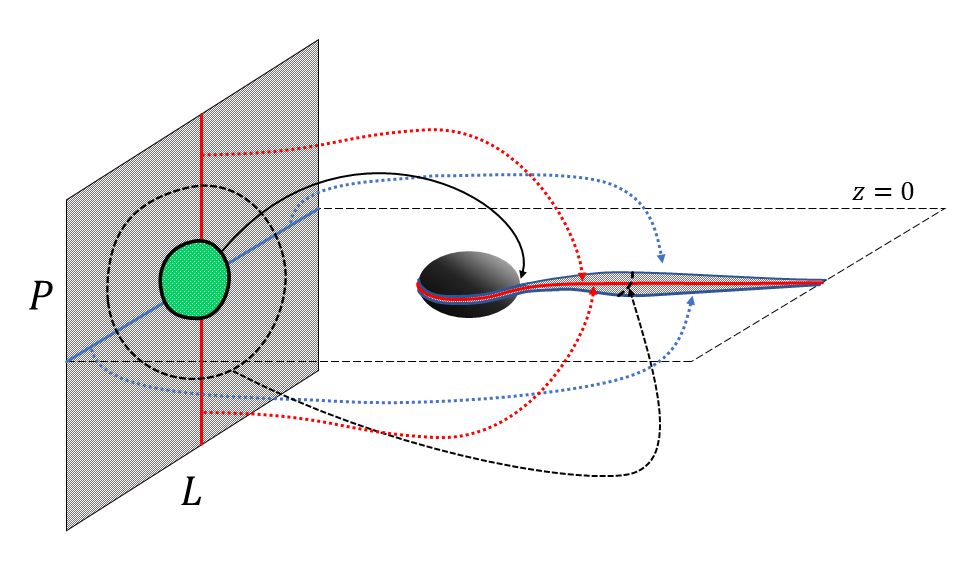}}
\caption{\small Sketch of how the crease-forming generators map the asymptotic $(L,P)$ plane to the crease strip (here widened for better visibility) on the equatorial plane in an orthogonal merger. This is like fig.~\ref{fig:thpi2xyz} but modified for clarity, with the same color coding of fig.~\ref{fig:Causticdisk}. The crease strip winds around the Kerr horizon an infinite number of times. The gray part of the plane, with $P\neq 0$, corresponds to generators that meet pairwise (with opposite values of $P$) at crossover points of the crease set. The horizontal blue lines along the $P=0$ axis on the plane map to the edges of the crease set; these are caustic points. The vertical red lines $L=0$ map to the center of the strip. The mapping is reflection-symmetric about the equatorial plane. \label{fig:CausticMap}}
\end{figure}

The way in which the null geodesic flow maps the gray region of the $(L,P)$  plane onto the crease strip is sketched in fig.~\ref{fig:CausticMap}. Generically, each pair of points with $(L,\pm P)$ outside the disk are mapped to one crossover point of the crease set. The points outside the disk along $P=0$  are mapped to the edges of the strip, which are caustic lines. A (reflection-symmetric) closed curve on the plane $(L,P)$ outside the critical curve is mapped onto a segment in the crease set made of crossover points. This mapping degenerates at the critical curve itself, $P=P_c(L)$, which is mapped to points on the equatorial circle of the Kerr horizon after winding an infinite number of times around it. 

When the rotation decreases the strip shrinks until, when $a=0$, the configuration becomes axisymmetric and the crease set becomes a line of caustics along the symmetry axis.

\begin{figure}[t!]
\centerline{\includegraphics[trim=0 40 0 40,clip,width=0.8\textwidth]{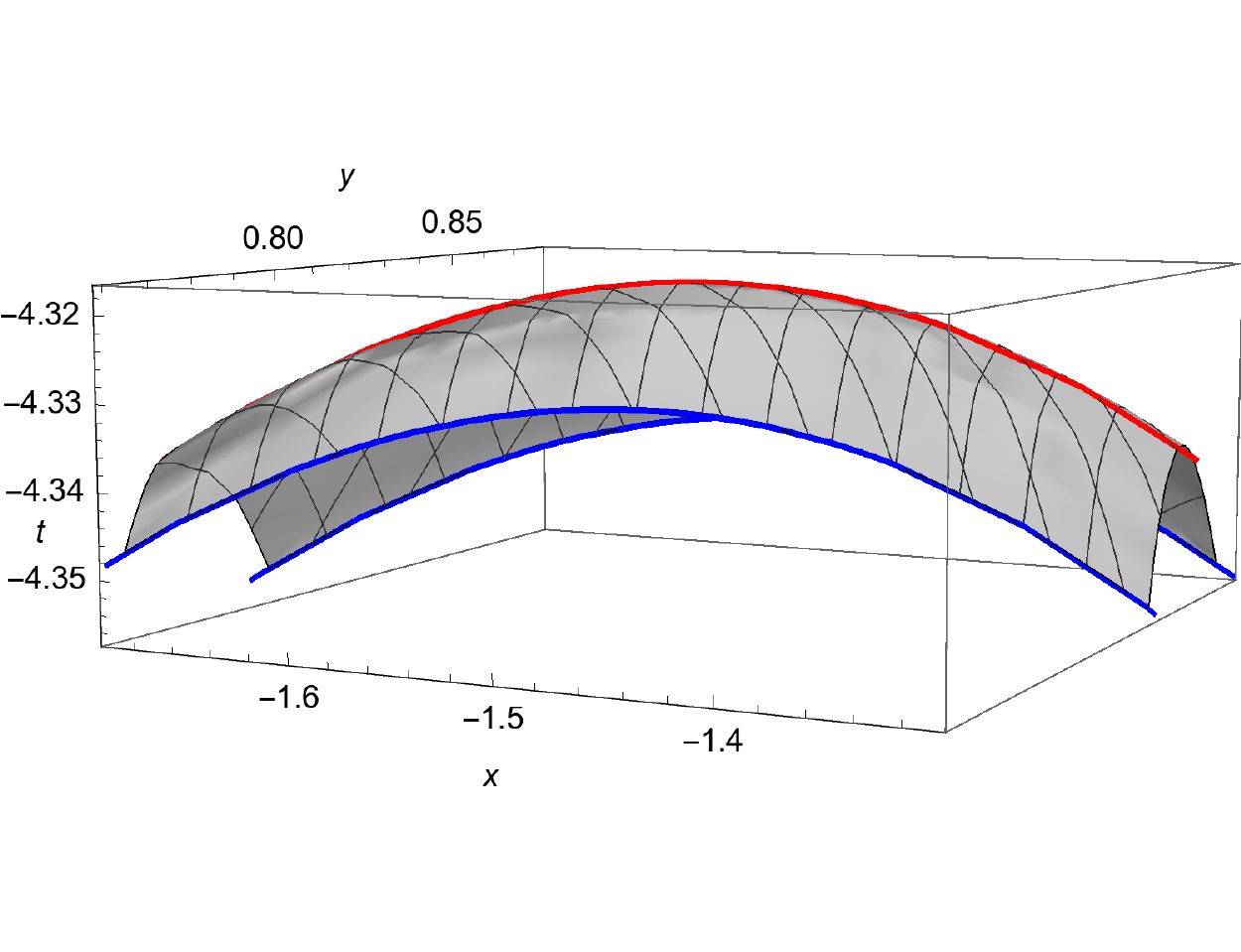}}
\caption[]{\small Projection on the $x,y,t$ coordinates of the crease set surface through which generators enter the event horizon. Gray points are crossovers of generators with $|P|\neq 0$. The central red line is formed by the crossover points of the generators with $L=0, |P|\neq 0$, whereas the blue lines are formed by the caustic points of the generators with  $P=0$, $L_0^+\leq L < \infty$ and $ -\infty< L \leq -L_0^-$. This surface is a zoomed version of the top portion of the red line in figure~\ref{fig:thpi2xyt}. When projected on the $(x,y)$ plane, it is a zoomed portion of the crease set strip in fig.~\ref{fig:CausticMap} near its thickest part. For this case $a=m$. \label{fig:causticsurftxy}}
\end{figure}
Let us now zoom in onto the spacetime region close to the merger point. The crease set is shown in fig.~\ref{fig:causticsurftxy} (recall that the crease set lies in the equatorial plane $z=0$). Imagine now slicing this surface in constant-time cuts. Then we reproduce the evolution of the horizon that we described earlier in sec.~\ref{subsec:constt} in which a transient toroidal phase occurs. At early times the surface is sliced into two segments shaped like $\subset\,\,\supset$. Then at the instant at which the blue lines (the caustic boundaries of the crease strip) reach a maximum in time, the segments close to form the hole in the horizon. Afterwards, the hole closes up, faster than the speed of light since its rim sweeps a spacelike surface in spacetime. In the preferred Killing time, the toroidal phase lasts for about a few percent of $m$ (more precisely, $0.027\,m$ in the $a=m$ example of fig.~\ref{fig:causticsurftxy}).

\section{Generic orientations}\label{sec:alphapi4}

Collisions with generic axis orientation, $0<\alpha<\pi/2$, turn out to be qualitatively similar to orthogonal collisions, but it is somewhat harder to produce clear visualizations of their properties. We have investigated the cases $\alpha=\pi/8$, $\pi/4$, and $3\pi/8$.

The numerical integration of the equations is unproblematic in any case, and the main issue in the construction is the identification of crossovers (caustics are zero-measure sets in the parameter space $(L,P_\alpha)$, hence numerically they are not easily seen directly). It turns out that, even if in the generic case the hypersurface does not have any symmetry, there is a discrete $\mathbb{Z}_2$ symmetry associated to the surface of crossovers which is inherited from the symmetry of the Kerr geometry. 

\begin{figure}[t!]
\centerline{\includegraphics[width=0.6\textwidth]{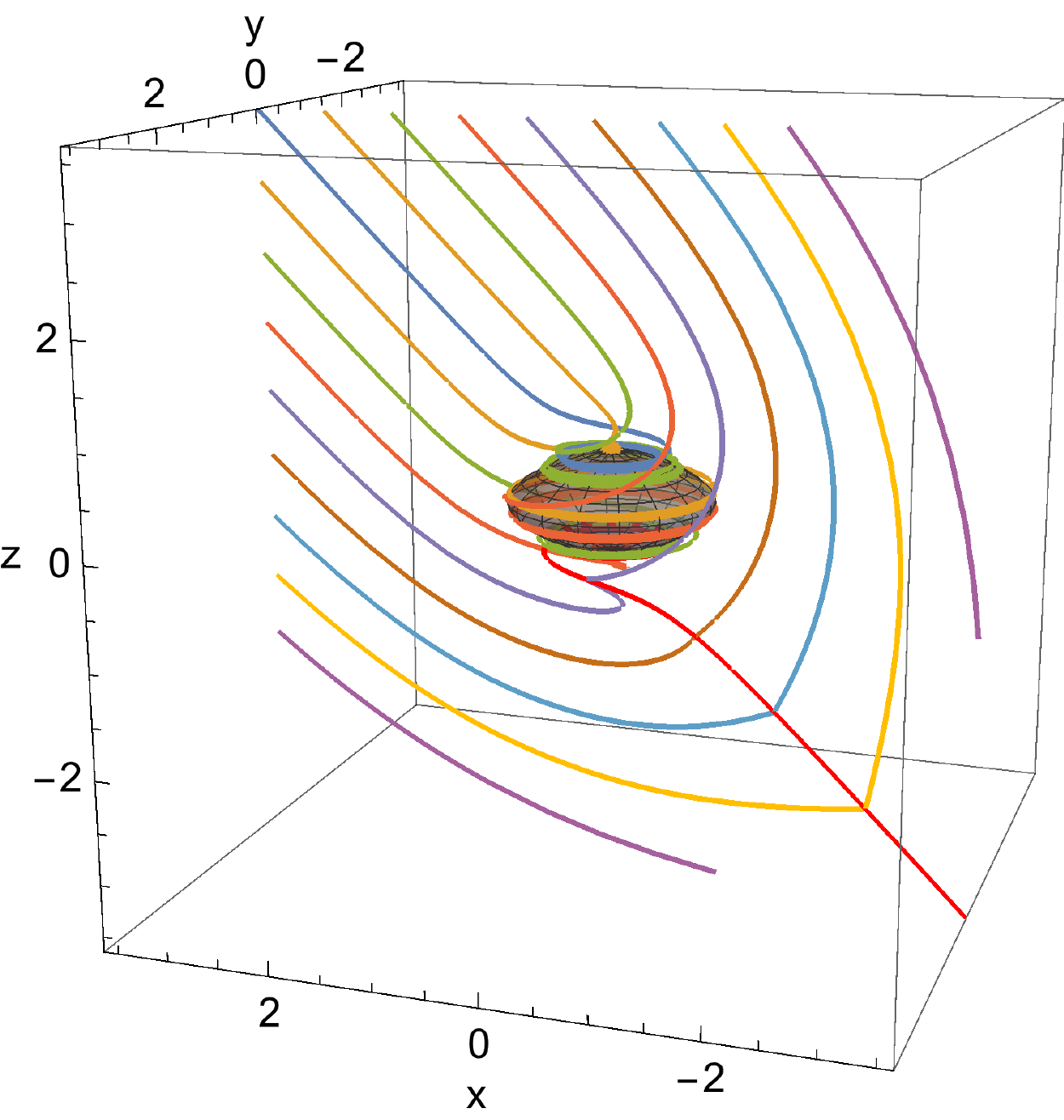}}
\caption[]{\small Generators of the event horizon for the collision with $\alpha=\pi/4$, projected on the $x,y,z$ space. We only show generators with $L_\alpha=0$. Those with the same magnitude of $P_\alpha$ but opposite signs have the same colour and intersect pairwise over the red line, which corresponds to the (very thin) crease strip. This figure represents an intermediate between figs.~\ref{fig:th0xyz} and \ref{fig:thpi2xyz}. \label{fig:thpi4xyzL0}}
\end{figure}

Specifically, we have found that the crossovers are generated by geodesics that meet pairwise, and that have the same value of $L_\alpha$ and equal magnitude of $P_\alpha$ but opposite signs. This can be seen in fig.~\ref{fig:thpi4xyzL0} where we illustrate the event horizon for these collisions in a spatial projection plot; it can be regarded as an intermediate situation between figs.~\ref{fig:th0xyz} and \ref{fig:thpi2xyz}. This symmetry is of great help in locating the crease set, since it implies that it lies on the surface of the cone defined by $\theta=\alpha+\pi$. The two previous cases correspond to the two degenerate limits in which the cone closes into a line ($\alpha=0$), or opens into a plane ($\alpha=\pi/2$). The crease strip narrows down monotonically with $\alpha$. For $\alpha=\pi/4$ it is approximately half the width of the $\alpha=\pi/2$ caustic. Other properties also appear to vary monotonically with $\alpha$.

Other than this, the mapping between the crease-forming region outside the creaseless disk in the asymptotic plane and the crease strip is qualitatively like in fig.~\ref{fig:CausticMap}.

\subsection{Influence of $\alpha$ on expansion factor and relaxation time} 

\begin{figure}[ht]
\centerline{\includegraphics[width=0.5\textwidth]{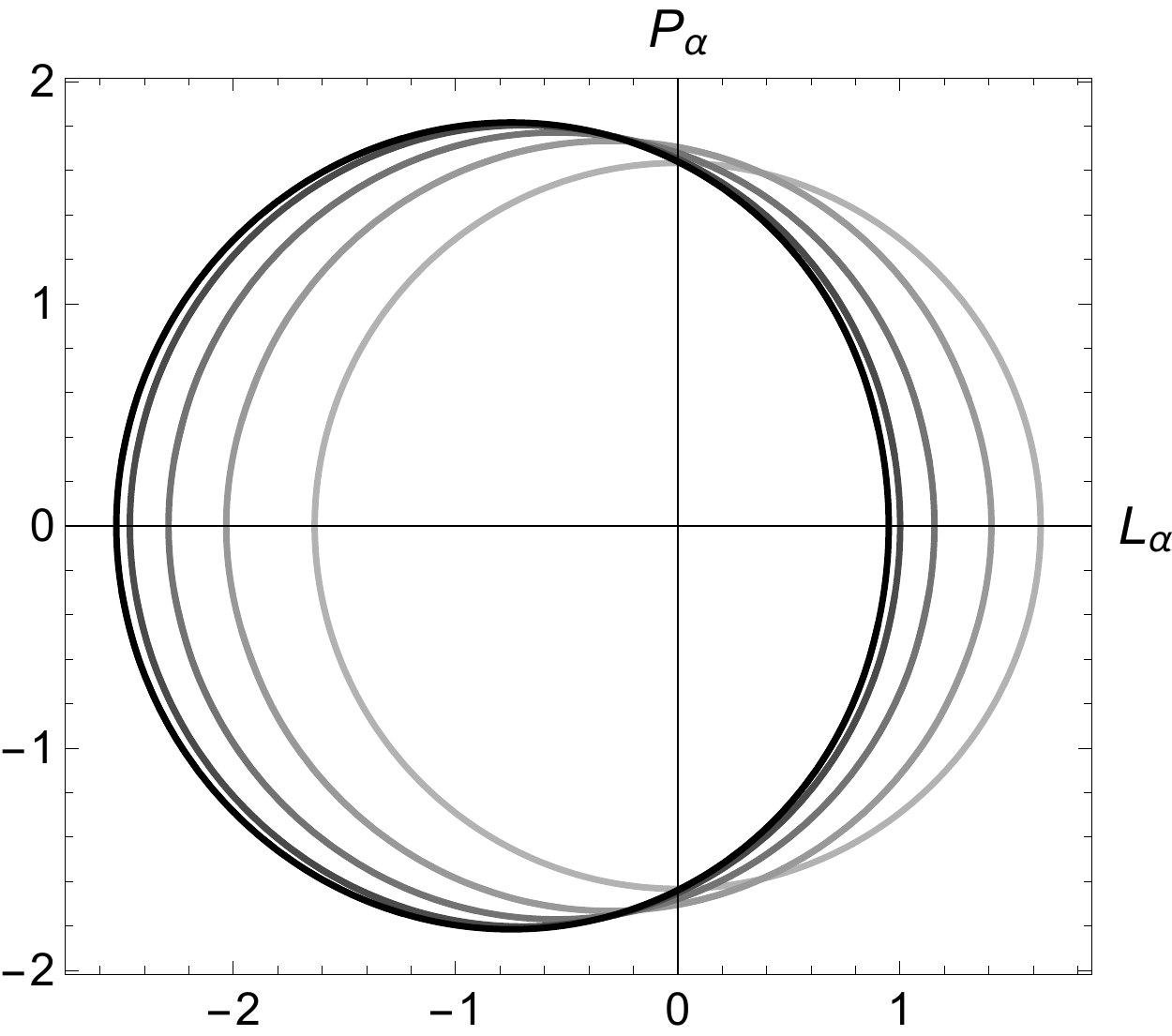}}
\caption[]{\small Creaseless disks for different orientations. From lighter to darker, $\alpha$ takes the values $0$, $\pi/8$, $\pi/4$, $3\pi/8$ and $\pi/2$. For all disks the mass is fixed and $a/m=1$. The increase in the area of the disk with $\alpha$ means that the further a merger is from axial symmetry, the more irreversible it is. \label{fig:creaselessdisks}}
\end{figure}

We estimate the area of the green disk in fig.~\ref{fig:Causticdisk}, which corresponds to an orthogonal merger in the extremal limit $a/m=1$, to be $\mathcal{A}_\mathrm{disk}\approx 39\, m^2$. In this case the creaseless generators of the horizon expand by a factor $F_\mathcal{A}\approx 1.57$. This is larger than the expansion factor $\approx 1.33$ in the aligned merger with $a/m=1$. In fact the expansion factor of the creaseless portion of the horizon grows monotonically with $\alpha$, as shown in fig.~\ref{fig:creaselessdisks}. The natural interpretation of this result is that the breakdown of the continuous symmetry enhances the irreversibility of the merger.

It is also possible to compare the relaxation times between mergers with different values of $\alpha$, since the integration constant $t_\infty=0$ is chosen independently of $\alpha$. We choose $t_*$ to be the instant where the hole in the toroidal horizon closes up, which corresponds to the maximum of the red line in fig.~\ref{fig:causticsurftxy}. 

\begin{figure}[t!]
\centerline{\includegraphics[width=0.5\textwidth]{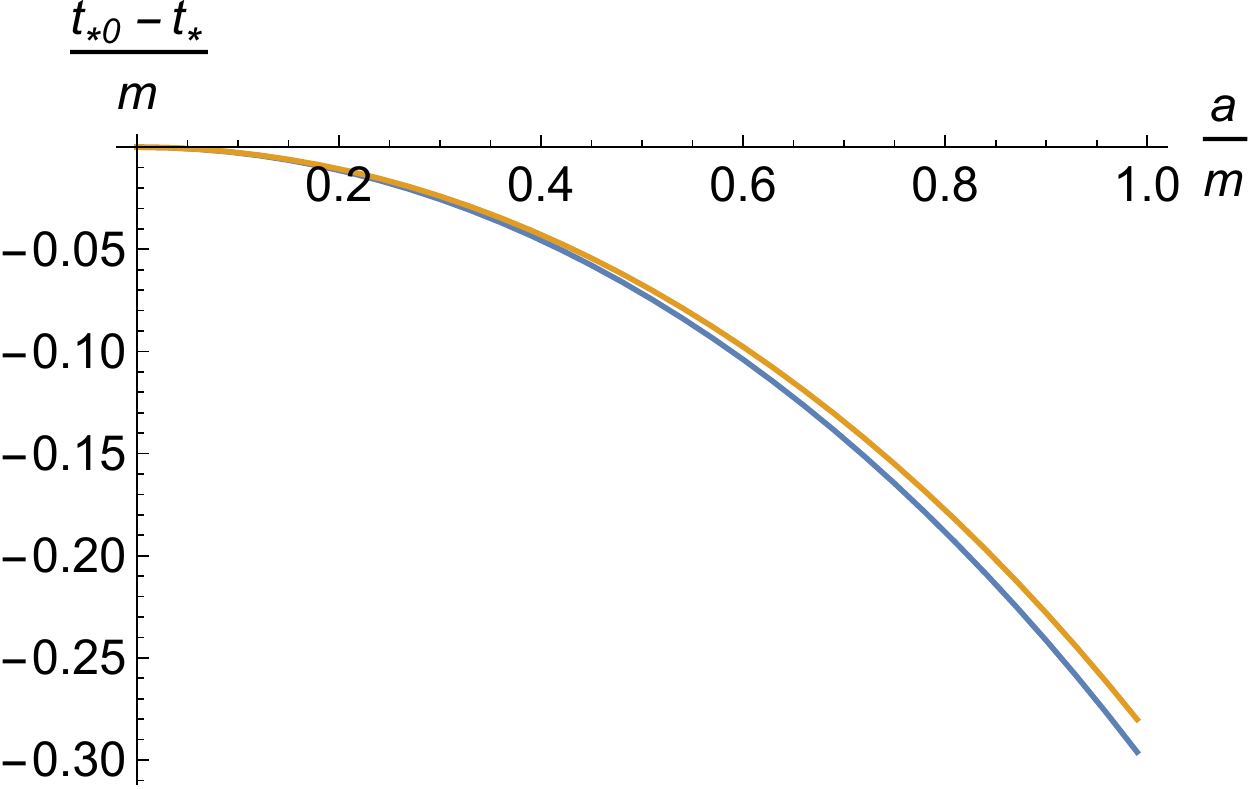}}
\caption[]{\small Relative relaxation times as a function of $a/m$ for the aligned ($\alpha=0$, blue line, same as fig.~\ref{fig:th0Dtvsa-fixedmass}) and orthogonal ($\alpha=\pi/2$, yellow line) mergers. For other $\alpha$ the results are almost the same as for $\alpha=\pi/2$. Relaxation times are then essentially independent of the orientation of the collision. \label{fig:trela}}
\end{figure}

In fig.~\ref{fig:trela} we show the results for fixed mass and different $a$, for $\alpha=0$ and $\alpha=\pi/2$. We see that, for all values of the rotation, the aligned, axisymmetric merger relaxes slightly more slowly than the orthogonal merger. Although the difference is very small, we do not find this result easily intuitive. We have also explored other orientations $(\alpha=\pi/8,\pi/4,3\pi/8)$ and the results coincide almost exactly with the values for $\alpha=\pi/2$. This indicates that the orientation of the collision affects the relaxation times very little. That is, relaxation times are mostly determined by the quantity of generators that the large black hole needs to accommodate, and not by how those generators are oriented. 

\section{Universal criticality in axisymmetric pinches}\label{sec:locpinch}

We have seen that in axisymmetric mergers the crease set is a line of caustics along the symmetry axis. On spacelike slices before the fusion, the caustics are conical points on each of the two sections of the horizon. The two black holes fuse together when the caustic cones in each of them degenerate to cusps and touch each other.

Reference~\cite{Emparan:2016ylg} studied the non-rotating case $a=0$, which always has axial symmetry, and focused on the critical behavior of the horizon around the pinch point at time $t_*$. It found evidence that: 
\begin{itemize}
\item[(i)] right before the pinch, the opening angle of the cone on the large black hole shrinks like $\sim \sqrt{t_*-t}$; 
\item[(ii)] right after the pinch, the width of the throat on the horizon grows linearly in time, \ie like $t-t_*$. 
\end{itemize}
Reference~\cite{Emparan:2016ylg} conjectured that this behavior is universal for all axisymmetric mergers where the crease set is a caustic line---including in particular black holes in all dimensions. 
Using the constructions of aligned mergers in sec.~\ref{sec:alpha0} we have checked numerically that, for any value of $a/m$, the behavior (i) is present for the cones not only of the large black holes but also of the small ones, and that (ii) also holds. 

We now proceed to prove that (i) and (ii) are exact, highly universal properties of axisymmetric mergers. To do so we use a simple local model of the horizon around the caustic line close to $t=t_*$. 

\subsection{Local model of pinching horizon}

Let us focus on very short scales close to the pinch point, so the local geometry is Minkowski space (below we will justify this approximation in more detail). Since we assume axial symmetry ($\alpha=0$) around the collision axis $z$, we write the local metric as
\beq
ds^2=-dt^2+dz^2+d\rho^2+\rho^2d\varphi^2\,.
\eeq
We now consider the following three-dimensional null congruence:
\beq\label{loceh}
t=t_*+\lambda\,,\qquad z=p_z\,,\qquad \rho=\lambda+\frac12 p_z^2\,,\qquad \varphi=c_\varphi\,.
\eeq
Here $\lambda$ is the affine parameter along the null rays, which are labeled by $p_z$ and $c_\varphi$. 

Let us now argue that this congruence is a local description of the event horizon close to the pinch point, such that the pinch occurs at $t=t_*$, on $\rho=z=0$.

First, one easily sees that the divergence of the tangent vector to the congruence, $\ell^\mu=\dot{x}^\mu=(1,0,1,0)$, is
\beq
\nabla_\mu \ell^\mu=\frac1\rho
\eeq
an therefore blows up at $\rho=0$. Thus the congruence has a caustic along the axis of symmetry $\rho=0$. Light rays coming from different angles $\varphi$ converge to focus there. 

\begin{figure}[t!]
\centerline{\includegraphics[width=\textwidth]{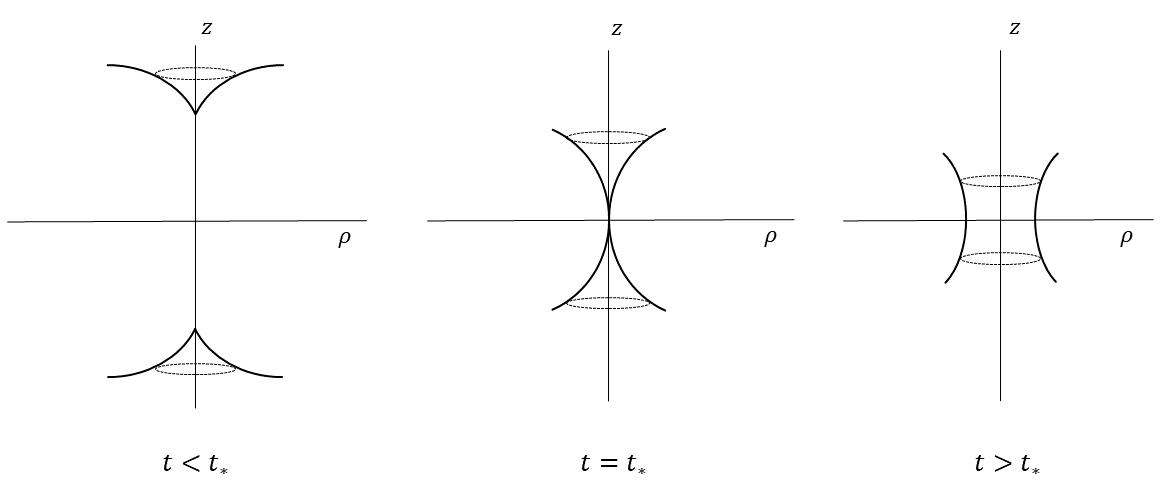}}
\caption[]{\small Constant-time slices of the local model \eqref{ehloc} of the event horizon around the pinch point. \label{fig:locmod}}
\end{figure}

Second, note that we can eliminate $\lambda$ to write \eqref{loceh} as the surface of revolution
\beq\label{ehloc}
t-t_*=\rho-\frac12 z^2\,,\qquad 0\leq\varphi<2\pi\,.
\eeq
The shape of constant-$t$ sections of this hypersurface is illustrated in fig.~\ref{fig:locmod}. This sequence describes a local pinch of the kind we have found in axisymmetric mergers. 

Third, one might be concerned that in the full merger configuration the pinch point is not far from the small black hole, so the approximation that the geometry around the pinch is flat may not hold. However, on general grounds the deviations from flatness are apparent only within scales $\sim m$ from the pinch, while this local model needs only extend out to distances $|z|\lesssim \sqrt{2m |t-t_*|}$. These are parametrically smaller than $m$, so our approximation holds well.\footnote{On dimensional grounds, we can write $\rho=\lambda+p_z^2/(2m)$ in \eqref{loceh}, and then the surface \eqref{ehloc} is $|z|=\sqrt{2m|t-t_*-\rho|}$.} This same argument allows us to neglect the effects of rotation of the horizon around the $z$ axis, \ie\ the congruence is locally non-rotating. 

Finally, observe that when $|z|\ll |t-t_*|,\,\rho$ the surface \eqref{ehloc} becomes a pair of null planes intersecting over $\rho=0$. In this limit the caustic is a straight line, which does not capture any of the time-evolution of the merger. The paraboloidal shape of \eqref{ehloc} is the first correction to this shape, due to the (extrinsic) curvature of the event horizon. The only assumption that we are making in requiring this form of the event horizon (other than smoothness) is the sign of this curvature: an inverted paraboloid shape might be possible, but it would yield a critical behavior inverted in time, contrary to what we have observed. 

We conclude that this local model captures the generic shape of the event horizon in axisymmetric pinches---not only in extreme mass ratio mergers, but indeed in any axisymmetric merger of two horizons, as long as the spatial slicing is chosen to conform to \eqref{ehloc}.

\subsection{Exact critical exponents}

Now it is easy to prove the critical behavior observed in \cite{Emparan:2016ylg}. Considering constant-time sections of the event horizon before the pinch, $t<t_*$, the (tangent of the) opening angle of the caustic cones is
\beq\label{ccone}
\left.\frac{d\rho}{dz}\right|_{\rho=0}=\pm \sqrt{2(t_*-t)}\,,
\eeq
so as $t\to t_*$ the cones close up to form singular cusps at a rate precisely of the form in point (i) above.

Next, at post-fusion times $t>t_*$ the horizon \eqref{ehloc} takes the shape of a paraboloidal throat, with a neck at $z=0$ of radius
\beq\label{cthroat}
\rho_\textrm{neck}=t-t_*\,,
\eeq
\ie\ it grows linearly in time, as described in (ii) above.

Clearly this construction can be extended to all dimensions: simply replace the $\varphi$ circles by spheres $\Omega_{D-3}$. It also applies regardless of whether the black hole is charged or not, or whether there is a cosmological constant of any sign, since the construction is only based on local properties of null hypersurfaces which apply on scales much smaller than any charge radius or cosmological radius. As in any critical phenomenon, the exponents $1/2$ and $1$ in \eqref{ccone} and \eqref{cthroat} are universal, while the specific multiplicative coefficients are not. The fact that these critical exponents are of mean-field type simply reflects that these horizons are smooth except at mild caustic singularities.

We may also write down local models of the geometry of the crease surfaces in non-axisymmetric fusions. These are spatial surfaces of the local form
\beq
t_*-t=k_{x} x^2+k_{y} y^2\,,
\eeq
where the constants $k_{x,y}$ characterize the principal curvatures of the crease surface near its limiting time $t_*$, and we assume both are positive\footnote{Bear in mind that this is a model of only the caustic surface, and the principal directions $x,\,y$ need not be aligned with any external directions.}. From this local model we see that the hole in the horizon closes up at an asymptotic rate $\dot{x},\,\dot{y}\sim (t_*-t)^{-1/2}$.

\section{Completeness of the construction}\label{sec:complete}

An important aspect of the construction that we have described---namely, drawing in the Kerr geometry null geodesics that at future null infinity asymptote to a specific null plane---is that it is complete. That is, this construction is sufficient to describe any fusion of neutral black holes in the extreme-mass ratio $m/M\to 0$. This means that it includes
\begin{itemize}
\item arbitrary spins of either black hole,
\item arbitrary relative orientations of the spins,
\item arbitrary infall trajectories,
\item arbitrary relative velocities. 
\end{itemize}

The argument that leads to this conclusion consists of two parts. The first is the uniqueness theorem of General Relativity: if the small black hole is uncharged, then its geometry must be that of the Kerr solution. Thus we only need to compute a particular family of null geodesics in the Kerr spacetime. 

The second part of the argument refers to the specification of this null plane that represents the large black hole horizon. Its choice in the Kerr spacetime (of the small black hole) amounts to specifying how the small and large black holes approach each other in the collision, \ie\ what their relative velocity and orientation are. Already our discussion in sec.~\ref{subsec:fixconst} has left no room for any physical parameter other than $a/m$ and $\alpha$ to characterize an arbitrary configuration, so we could conclude here, but let us argue the generality of the construction from a slightly different perspective.

Begin with the simplest case where the small black hole (ignoring its spin for now) plunges into the large one in a head-on trajectory, that is, orthogonal to the large planar horizon. Any velocity along this direction amounts to a simple boost along the collision axis. But this does not affect the asymptotic null plane, which is invariant under boosts along this axis, so these velocities do not modify the construction. 

Next, consider that there is a relative motion in a direction parallel to the large black hole. This motion could be due to the angular velocity of the large black hole, or to a skewed collision angle---the two situations are indistinguishable in the limit $m/M\to 0$ that we work in. It was argued in \cite{Emparan:2016ylg} that a relative velocity $\mathbf{v}$ between the two black holes in a direction parallel to the large black hole, can be traded for a rotation of the asymptotic null plane, on the spatial plane spanned by $\mathbf{v}$ and the normal $\mathbf{n}$ to the null plane, with angle $\arcsin |\mathbf{v}|$. This brings us back to a configuration of the type we have described, generically with different integration constants $\theta_\infty$ and $\varphi_\infty$.\footnote{The angle $\alpha$ is not invariant under these transformations, so we define it in the frame where the relative transverse velocity $\mathbf{v}$ vanishes.} We conclude that we do not need to consider different final null planes to account for any relative translational motion between the two colliding black holes. 

Thus, in the limit $M\to \infty$ the only non-trivial relative motion between the two black holes is the rotation of the small black hole. This is parametrized by $a/m$ (the mass scale $m$ can always be set to one). 

Hence, with $a/m\in [0,1]$ and $\alpha\in [0,\pi/2]$ the method we have presented suffices to accurately describe all the mergers between any astrophysical black holes in the Universe, as long as their mass ratio is small enough.

\section{Final remarks}\label{sec:concl}

The event horizon is, by definition, not observable from far away. One may even question the practical, astronomical observability, of the evolution of a surface close to it, but outside it, in a highly dynamical situation such as a merger. Moreover, the event horizon is well-known to behave in peculiar ways. For instance, its absolute finality---nothing \textit{ever} escapes from it---is the reason that, as we have witnessed in our constructions, the event horizon does not behave in a future-causal manner, appearing instead to eerily anticipate which objects will eventually be behind it and so unable to escape away. 

All of this, however, does not imply that the event horizon is an unphysical contrivance. It is a null surface that does exist in the full spacetime geometry, and could ideally be determined if one knew the history of the latter in close enough detail. Thus we regard the event horizon as real and physical.  

The relevance of our specific construction of the event horizon in the limit $m/M\to 0$ was already discussed in \cite{Emparan:2016ylg}. For the convenience of readers, we summarize here the main points. First, this construction provides a simple setup where the features of a highly dynamical event horizon can be studied in great detail without the use of supercomputer simulations. Second, it provides a benchmark for numerical calculations in a regime where, due to the large disparity between scales, these techniques become increasingly hard. And, third, it is the leading-order solution for the collision of two black holes in a perturbative expansion in the small parameter $m/M$, which can then be taken to higher orders via a matched-asymptotic expansion. In this scheme, our calculation should be matched to the analyses in \cite{Hamerly:2010cr,Hussain:2017ihw} of the event horizon to leading order in $m/M$, where now one takes $M$ to be fixed while $m\to 0$. Already one can see that the features of the event horizon coincide between these studies and ours, in the regime of intermediate length scales, $m\ll r\ll M$, where both approaches should be applicable. It will be interesting to develop these matched-asymptotic expansions further. In this article we have provided the leading order solution. Higher orders in the expansion will bring back in the gravitational radiation produced in the merger.

\section*{Acknowledgments}
We are grateful to Ivan Booth and Barak Kol for discussions. RE acknowledges the hospitality of Centro de Ciencias de Benasque ``Pedro Pascual'' during the programme ``Gravity - New perspectives from strings and higher dimensions'' in July 2017. The work of R. E. and M. M. was supported by grants FPA2013-46570-C2-2-P and FPA2016-76005-C2-2-P from MINECO (Spain); grant 2009-SGR-168 from AGAUR (Catalunya); and Advanced Grant No. GravBHs-692951 from European Research Council. The work of M.Z. was supported by Starting Grant No. HoloLHC-306605 from European Research Council, and Program No. IF/00729/2015 from FCT (Portugal).


\addcontentsline{toc}{section}{Appendix}
\appendix
\section{Asymptotic solution}\label{app:asympsol}

We begin by rewriting equation~\eqref{eq:dotrdottheta} as
\begin{equation}
\sigma\frac{dr}{\sqrt{\left(a^2-a L+r^2\right)^2-\Delta  \left((a-L)^2+Q\right)}}=\frac{d\theta}{\sqrt{Q+a^2 \cos ^2\theta-L^2 \cot ^2 \theta }}\,.
\end{equation}
Then we expand the left hand side around $r=\infty$, and the right hand side around $\theta=\alpha$, integrate each of the sides of the equation, and invert the series to obtain 
\begin{align}
\begin{split}
r(\theta)=&-\frac{P_\alpha}{\theta -\alpha }+\frac{\sin\alpha \left(a^2 \cos\alpha-L^2 \cot\alpha \csc ^3\alpha\right)}{2 P_\alpha}+\mathcal{O}\left(\theta -\alpha\right),
\end{split}\label{eq:r_theta}\\
\begin{split}
\theta(r)=&\,\alpha -\frac{P_\alpha}{r}+\frac{\sin\alpha \left(L^2 \cot\alpha \csc ^3\alpha-a^2 \cos\alpha\right)}{2 r^2}+\mathcal{O}\left(\frac{1}{r^3}\right).
\end{split}\label{eq:theta_r}
\end{align}
($P_\alpha$ was introduced in \eqref{Palpha}).
These can now be plugged back into \eqref{eq:dotrdottheta} to solve for $\lambda(r)$ and $\lambda(\theta)$ as series expansions in $1/r$ and $1/(\theta-\alpha)$, respectively. Inverting these series we find $r(\lambda)$ and $\theta(\lambda)$. Using these solutions it is now straightforward to integrate the remaining equations in \eqref{eq:geo}. 
The resulting set of solutions is
\begin{subequations}
	\allowdisplaybreaks
	\label{eq:IC}
	\begin{align}
	t(\lambda)&=\,\lambda+2m \, \log \lambda  -\frac{4 m^2}{\lambda } \nonumber\\
	&\quad+\frac{m \left(3 a^2 \cos ^2\alpha-a^2+2 a L+L^2 \csc ^2\alpha-8 m^2+P_{\alpha }^2\right)}{2 \lambda ^2}+\mathcal{O}\left(\frac{1}{\lambda^3}\right) \,,
	\\
	r(\lambda) & =\lambda +\frac{P_\alpha^2-a^2\sin^2\alpha +L^2 \csc^2\alpha}{2 \lambda } 
\nonumber\\
	& \quad -\frac{m (L \csc \alpha -a \sin \alpha )^2-a^2 P_{\alpha } \sin 2 \alpha + m\,P_{\alpha }^2}{2 \lambda ^2} +\mathcal{O}\left(\frac{1}{\lambda^3}\right) \,,
	\\
	\theta(\lambda)&=\,\alpha -\frac{ P_\alpha}{\lambda }+\frac{2 L^2 \cot \alpha  \csc ^2\alpha -a^2 \sin 2 \alpha }{4 \lambda ^2}+\mathcal{O}\left(\frac{1}{\lambda^3}\right)  \,, \\
	\varphi(\lambda)&=-\frac{L \csc ^2\alpha}{\lambda } - \frac{  P_\alpha\,L\, \cot\alpha \csc ^2\alpha +a \, m}{\lambda ^2}+\mathcal{O}\left(\frac{1}{\lambda^3}\right)  \,,
	\\
	p_r(\lambda)&=\,1+\frac{2 m}{\lambda }-\frac{P_\alpha^2+a^2\sin^2\alpha+L^2\csc^2\alpha-8 m^2}{2 \lambda ^2}+\mathcal{O}\left(\frac{1}{\lambda^3}\right) \,,
	\\
	p_\theta(\lambda)&=\, P_\alpha +\frac{a^2 \sin\alpha \cos\alpha-L^2 \cot\alpha \csc ^2\alpha}{\lambda }\nonumber\\
	& \quad-\frac{P_{\alpha } \left(a^2 \cos 2\alpha +L^2 (\cos 2\alpha +2) \csc ^4\alpha \right)}{2 \lambda ^2}+\mathcal{O}\left(\frac{1}{\lambda^3}\right) \,.
	\end{align}
\end{subequations}


\end{document}